\documentclass[twocolumn,english,showpacs,preprintnumbers,amsmath,amssymb,floatfix,nofootinbib]{revtex4-1}

\usepackage[T1]{fontenc}
\usepackage[latin9]{inputenc}
\usepackage{color}
\usepackage{array}
\usepackage{amstext}
\usepackage{graphicx}
\usepackage{esint}
\usepackage{rotating}
\usepackage{slashed}
\usepackage{siunitx}
\usepackage[font=small,labelfont=bf]{caption}
\usepackage{soul,xcolor}

\usepackage{xcolor}

\usepackage[colorlinks = true,
linkcolor = magenta,
urlcolor  = blue,
citecolor = red,
anchorcolor = blue]{hyperref}

\usepackage{ulem}

\makeatletter

\begin{document}

\setstcolor{blue}

\preprint {MITP/20-043}

%
%
\title{ 
Simultaneous extraction of fragmentation functions 
\\
of light charged hadrons with mass corrections 
} 
%
%

\author{Maryam Soleymaninia$^{1}$}
\email{Maryam\_Soleymaninia@ipm.ir}

\author{Muhammad Goharipour$^{1}$}
\email{Muhammad.Goharipour@ipm.ir}

\author{Hamzeh Khanpour$^{2,1}$}
\email{Hamzeh.Khanpour@cern.ch}

\author{Hubert Spiesberger$^{3}$}
\email{spiesber@uni-mainz.de}

\affiliation {
$^{1}$School of Particles and Accelerators, 
Institute for Research in Fundamental Sciences (IPM), 
P.O.Box 19395-5531, Tehran, Iran 
\\
$^{2}$Department of Physics, University of Science and 
Technology of Mazandaran, P.O.Box 48518-78195, 
Behshahr, Iran 
\\
$^{3}$PRISMA{\color{red}$^{+}$} Cluster of Excellence, 
Institut f\"ur Physik, Johannes-Gutenberg-Universit\"at, 
Staudinger Weg 7, D-55099 Mainz, Germany
}

\date{\today}

%
\begin{abstract}

Achieving the highest possible precision for theoretical 
predictions at the present and future high-energy lepton 
and hadron colliders requires a precise determination of 
fragmentation functions (FFs) of light and heavy charged 
hadrons from a global QCD analysis with great accuracy. 
We describe a simultaneous determination of unpolarized 
FFs of charged pions, charged kaons and protons/antiprotons 
from single-inclusive hadron production in electron-positron 
annihilation (SIA) data at next-to-leading order and 
next-to-next-to-leading order accuracy in perturbative 
QCD. A new set of FFs, called {\tt SGKS20}, is presented.  
We include data for identified light charged hadrons ($\pi^\pm, 
K^\pm$ and $p/\bar{p}$) as well as for unidentified light 
charged hadrons, $h^\pm$ and show that these 
data have a significant impact on both size and uncertainties 
of the fragmentation functions. We examine the inclusion of 
higher-order perturbative QCD corrections and finite-mass 
effects. We compare the new {\tt SGKS20} FFs with other 
recent FFs available in the literature and find in general 
reasonable agreement, but also important differences for some 
parton species. We show that theoretical predictions obtained 
from our new FFs are in very good agreement with the analyzed 
SIA data, especially at small values of $z$. The {\tt SGKS20} 
FF sets presented in this work are available via the {\tt LHAPDF} 
interface.

\end{abstract}
%


\maketitle
\tableofcontents{}

%
\section{Introduction}\label{sec:introduction}
%

The two non-perturbative elements in theoretical high energy cross 
sections of hard scattering processes are the parton distribution 
functions (PDFs) and the collinear unpolarized fragmentation functions 
(FFs)~\cite{Bertone:2018ecm,Soleymaninia:2019sjo,Bertone:2017tyb,Soleymaninia:2017xhc,
Soleymaninia:2018uiv,Soleymaninia:2013cxa,Epele:2018ewr,
deFlorian:2007aj,deFlorian:2007ekg,Boroun:2016zql,Ethier:2017zbq,Albino:2008fy}. 
The factorization theorem of Quantum Chromodynamics (QCD) tells us 
that these are universal and their evolution can be calculated from 
perturbative QCD. A precise determination of FFs is crucial for studies 
of the strong interaction in high energy scattering processes. FFs 
describe how high energy colored partons produced in the hard 
interactions are turned into the hadrons measured and identified 
in an experiment. As is the case for PDFs, FFs need to be determined 
through a QCD analysis of high-energy experimental data due to their 
non-perturbative nature. Currently, several experimental measurements 
from different processes are available which can be used for the 
determination of FFs. Hadron production in single-inclusive $e^+e^-$ 
annihilation (SIA) provides the main information on FFs, but 
measurements from semi-inclusive deep inelastic scattering (SIDIS) 
and from proton-(anti)proton collisions at hadron colliders can also be 
used to determine well-constrained FFs. SIDIS and proton-proton 
collisions are particularly important for a complete flavour 
decomposition of FFs into quark and anti-quark components. However, 
among these high-energy processes, SIA is the cleanest process and 
the interpretation of it does not require a simultaneous knowledge 
of PDFs. 

There have been several analyses aiming to extract FFs of the lightest 
charged hadrons $\pi^{\pm}$, $K^{\pm}$ and 
$p/\bar{p}$~\cite{Soleymaninia:2019sjo,Bertone:2017tyb,Epele:2018ewr,deFlorian:2014xna,deFlorian:2017lwf,Sato:2016wqj,Anderle:2015lqa,Anderle:2016czy,Albino:2006wz,Kniehl:2000fe}. 
The most important experimental information for determining the 
FFs comes from SIA data and most of the recent analyses have considered  
only these data to determine FFs up to next-to-next-to-leading order 
(NNLO) in perturbative QCD. For the case of charged pion, kaon and 
proton/antiproton analyses which include SIDIS and $pp$ data, we refer 
to Refs.~\cite{deFlorian:2014xna,deFlorian:2017lwf,Albino:2008fy}. 

The analyses performed so far for extracting $\pi^{\pm}$, $K^{\pm}$ 
and $p/\bar{p}$ FFs differ in various aspects, such as the experimental 
data included, the QCD perturbative order, the phenomenological 
framework, the error calculation procedure, and so on. In particular 
we note that up to now, it was customary to analyze the light charged 
hadron data independently from each other, i.e.\ the extraction of FFs 
for one type of hadron was solely performed through the analysis of  
production data for that type of hadron. In contrast, in our most 
recent study~\cite{Soleymaninia:2019sjo}, we have shown, for the first 
time, that a simultaneous analysis of pion and unidentified light 
charged hadron data for extracting pion FFs is also possible and leads 
to a reduction in the uncertainties of the extracted pion FFs. 

The main goal of the following study, referred to as {\tt SGKS20} FFs, 
is to revisit our previous analysis~\cite{Soleymaninia:2019sjo} and 
extract $\pi^\pm$, $K^\pm$ and $p/\bar{p}$ FFs simultaneously by 
including all available SIA data for pion, kaon, proton production  
along with data for unidentified light charged hadrons $h^{\pm}$. 
We perform a QCD analysis at next-to-leading order (NLO) as well as 
at next-to-next-to-leading order (NNLO). Moreover, in the present 
analysis, we also study hadron mass corrections. We find that these  
corrections are important at small $z$, the ratio of momentum 
transferred from the parton to the observed hadron, and at low values 
of center-of-mass energy $\sqrt{s}$. Since the contribution of 
unidentified light charged hadrons $h^\pm$ is mostly related to the 
pion, kaon and proton, we show that the extraction of $\pi^{\pm}$, 
$K^{\pm}$ and $p/\bar{p}$ FFs in a simultaneous analysis of identified 
and unidentified light charged hadron production data and including the 
hadron mass corrections significantly improves the fit quality and 
leads to well-constrained FFs.

This article is organized in the following manner. In 
Sec.~\ref{sec:data_selection} we present
the SIA data used in our NLO and NNLO FFs analyses, along with their 
corresponding observables and the kinematic cuts we impose on the data. 
Then, in Sec.~\ref{sec:QCD_analysis} we discuss the theoretical details 
of the {\tt SGKS20} FFs determination of $\pi^\pm$, $K^\pm$ and 
$p/\bar{p}$ FFs, including the parameterizations and the evolution of 
FFs. Our assumptions and the hadron mass corrections are discussed in 
this section as well.  
Sec.~\ref{sec:minimizations} deals with the method of $\chi^2$ 
minimization and estimation of the {\tt SGKS20} FFs uncertainties. 
Considering the best fit parameters, the main  
results of this study are presented in Sec.~\ref{sec:Results}. 
We first turn to discuss the {\tt SGKS20} FFs sets. Then, we compare 
our best fit obtained for pion, kaon and proton/antiproton FFs at NNLO 
with other results in the literature. We also present a detailed 
comparison between all analyzed SIA data and the corresponding 
theoretical predictions obtained using the {\tt SGKS20} FFs. Finally, 
in Sec.~\ref{sec:conclusion} we present our summary and conclusions. 
We also outline in this section some possible future developments.

\begin{table*}[htb]
\renewcommand{\arraystretch}{2}
\centering 	\scriptsize
\begin{tabular}{lccccr}				\hline
Experiment        ~&~  $\sqrt{s}$ ~&~  $\frac{\chi^2}{N_\text{pts.}} (\pi^\pm)$    ~&~  $\frac{\chi^2}{N_\text{pts.}} (K^\pm)$ ~&~ $\frac{\chi^2}{N_\text{pts.}}(p/\bar{p})$ ~&~ $\frac{\chi ^2}{N_\text{pts.}}(h^\pm)$ 
\rule[-3mm]{0mm}{8mm}
\\
				\hline \hline
				{\tt BELLE}\cite{Leitgab:2013qh}   & 10.52  & 0.467& 0.966&---& --- \\
				{\tt BABAR}~\cite{Lees:2013rqd} &   10.54 & 1.793 & 2.838&1.017&--- \\
				{\tt TASSO12}~\cite{Brandelik:1980iy}   & 12 & 1.154 &0.930&0.648&--- \\
				{\tt TASSO14}~\cite{Althoff:1982dh,Braunschweig:1990yd}  & 14 & 1.202& 1.447 &2.237&0.607 \\       					
				{\tt TASSO22}~\cite{Althoff:1982dh,Braunschweig:1990yd}   & 22 & 2.461& 2.472&1.969&0.628\\
				{\tt TPC}~\cite{Aihara:1988su}  & 29 & 0.601 &0.664&4.419& 0.636\\
				{\tt TASSO30}~\cite{Brandelik:1980iy}   & 30 & --- &---&1.239&--- \\
				{\tt TASSO34}~\cite{Braunschweig:1988hv} & 34 & 1.265&  0.136&1.704&---\\
				{\tt TASSO35}~\cite{Braunschweig:1990yd}& 34 & ---&  ---&---&1.165 \\
				{\tt TASSO44}~\cite{Braunschweig:1988hv,Braunschweig:1990yd}& 44 & 2.052 &---&--- &0.770\\
				{\tt ALEPH}~\cite{Buskulic:1994ft,Buskulic:1995aw}  & 91.2 & 1.876 &0.797&2.248&0.814\\
				{\tt DELPHI} (incl.)~\cite{Abreu:1998vq}   & 91.2 & 1.274 &0.731&0.559&0.537\\
				{\tt DELPHI} ($uds$ tag)~\cite{Abreu:1998vq}  & 91.2 & 0.813 &1.062&0.671&0.378 \\
				{\tt DELPHI} ($b$ tag)~\cite{Abreu:1998vq}  &91.2 & 0.928  & 0.632 & 0.817&0.374\\
				{\tt OPAL} (incl.)~\cite{Akers:1994ez,Ackerstaff:1998hz} & 91.2 & 1.455& 0.879&---&0.682\\
				{\tt OPAL} ($uds$ tag)~\cite{Akers:1994ez,Ackerstaff:1998hz}   & 91.2 & ---  &---&---&0.554\\
				{\tt OPAL} ($c$ tag)~\cite{Akers:1994ez,Ackerstaff:1998hz}  &91.2 & ---  & ---&---&0.619 \\
				{\tt OPAL} ($b$ tag)~\cite{Akers:1994ez,Ackerstaff:1998hz}  & 91.2 & ---  &---&---&0.232\\
				{\tt SLD} (incl.)~\cite{Abe:2003iy}   & 91.2 & 1.865&0.578&0.824&0.307\\
				{\tt SLD} ($uds$ tag)~\cite{Abe:2003iy}   &91.2 & 1.602 & 2.045& 1.690&0.669 \\
				{\tt SLD} ($c$ tag)~\cite{Abe:2003iy} & 91.2 & 0.880  & 1.087&2.905&0.592 \\
				{\tt SLD} ($b$ tag)~\cite{Abe:2003iy} &91.2 & 0.702 &1.214&2.888 &0.170\\ 				
\hline \hline
Total $\chi^2/{\rm d.o.f.}$ &  &  & 1685.057/1438 = 1.171  \\
\hline \hline	
\end{tabular}
\caption{ \small 
  The list of input data sets for $\pi^\pm$, $K^\pm$, $p/\bar{p}$, 
  and $h^\pm$ production included in the present analysis. For each 
  data set, we have indicated the corresponding reference and the 
  center-of-mass energy $\sqrt{s}$. In the last four columns 
  we show the value of $\chi^2/{N_\text{pts.}}$ resulting from the 
  FF fit at NLO order. The total value of $\chi^2/{{\rm d.o.f.}}$ 
  is shown at the bottom of the table.  }
\label{tab:datasetsNLO}
\end{table*}

\begin{table*}[htb]
\renewcommand{\arraystretch}{2}
\centering 	\scriptsize
\begin{tabular}{lccccr}				\hline
Experiment        ~&~  $\sqrt{s}$ ~&~  $\frac{\chi^2}{N_\text{pts.}} (\pi^\pm)$    ~&~  $\frac{\chi^2}{N_\text{pts.}} (K^\pm)$ ~&~ $\frac{\chi^2}{N_\text{pts.}}(p/\bar{p})$ ~&~ $\frac{\chi ^2}{N_\text{pts.}}(h^\pm)$ 
\rule[-3mm]{0mm}{8mm}
\\
				\hline \hline
				{\tt BELLE}~\cite{Leitgab:2013qh}   & 10.52  & 0.295& 0.993&---& ---\\
				{\tt BABAR}~\cite{Lees:2013rqd} &   10.54 & 1.504& 2.503&0.234&---\\
				{\tt TASSO12}~\cite{Brandelik:1980iy}   & 12 & 1.135 &0.933&0.669&---\\
				{\tt TASSO14}~\cite{Althoff:1982dh,Braunschweig:1990yd}  & 14 & 1.194& 1.392 &2.166&0.627 \\       					
				{\tt TASSO22}~\cite{Althoff:1982dh,Braunschweig:1990yd}   & 22 & 2.348& 2.580&1.920&0.697 \\
				{\tt TPC}~\cite{Aihara:1988su}  & 29 & 1.099  &0.519&4.814& 0.438\\
				{\tt TASSO30}~\cite{Brandelik:1980iy}   & 30 & ---&---&1.339&---\\
				{\tt TASSO34}~\cite{Braunschweig:1988hv} & 34 & 1.136&  0.175&1.496&---\\
				{\tt TASSO35}~\cite{Braunschweig:1990yd}& 34 & ---&  ---&---&1.362 \\
				{\tt TASSO44}~\cite{Braunschweig:1988hv,Braunschweig:1990yd}& 44 & 2.129 &---&--- &0.799 \\
				{\tt ALEPH}~\cite{Buskulic:1994ft,Buskulic:1995aw}  & 91.2 & 1.362 &0.747&0.991&0.738 \\
				{\tt DELPHI} (incl.)~\cite{Abreu:1998vq}   & 91.2 & 1.471 &0.684&0.541 &0.508\\
				{\tt DELPHI} ($uds$ tag)~\cite{Abreu:1998vq}  & 91.2 & 0.991  &1.050&0.578&0.413 \\
				{\tt DELPHI} ($b$ tag)~\cite{Abreu:1998vq}  &91.2 & 0.850  & 0.651 & 1.537&0.295\\
				{\tt OPAL} (incl.)~\cite{Akers:1994ez,Ackerstaff:1998hz} & 91.2 & 1.380& 1.126&---&0.780\\
				{\tt OPAL} ($uds$ tag)~\cite{Akers:1994ez,Ackerstaff:1998hz}   & 91.2 & ---  &---&---&0.552 \\
				{\tt OPAL} ($c$ tag)~\cite{Akers:1994ez,Ackerstaff:1998hz}  &91.2 & --- & ---&---&0.624 \\
				{\tt OPAL} ($b$ tag)~\cite{Akers:1994ez,Ackerstaff:1998hz}  & 91.2 & --- &---&---&0.175\\
				{\tt SLD} (incl.)~\cite{Abe:2003iy}   & 91.2 & 1.181 &0.549&0.831&0.289 \\
				{\tt SLD} ($uds$ tag)~\cite{Abe:2003iy}   &91.2 & 1.186 & 2.065& 1.197&0.604 \\
				{\tt SLD} ($c$ tag)~\cite{Abe:2003iy} & 91.2 & 0.818  & 0.992&3.661&0.617 \\
				{\tt SLD} ($b$ tag)~\cite{Abe:2003iy} &91.2 & 0.667  &1.282&2.664 &0.140\\ 				
\hline \hline
Total $\chi^2/{\rm d.o.f.}$ &  &  &  1558.169/1438 =  1.083  \\
\hline \hline	
\end{tabular}
\caption{ \small 
  Same as Table.~\ref{tab:datasetsNLO} but for the {\tt SGKS20} FFs 
  fit at NNLO.
}
\label{tab:datasetsNNLO}
\end{table*}

\section{ Experimental observables } \label{sec:data_selection}

The SIA processes have provided us with a wealth of high-precision 
experimental data carrying information about how partons fragment 
into a low-mass charged hadron. In this section, we provide details 
of the experimental measurements used as input for the determination 
of the {\tt SGKS20} FFs along with the corresponding observables and 
kinematic cuts applied. The simultaneous determination of light 
charged hadron FFs presented in this work is based on comprehensive 
data sets from electron-positron annihilation into a single identified 
and unidentified hadron. In addition to the inclusive measurements, 
the data set entering the {\tt SGKS20} analysis also includes 
flavor-tagged measurements.

We note that SIA data are particularly clean, however, they provide 
only a limited sensitivity to the flavor separation of different 
light quark FFs. In addition, it is known that the gluon FF is poorly 
constrained by the total SIA cross section measurements. Hence, in 
order to improve the discrimination between different quark and 
antiquark flavors, one would have to include SIDIS and hadron 
collider observables. This is, however, beyond the scope of the 
present work. 

In our analysis of $\pi^{\pm}$, $K^{\pm}$, 
$p/\bar{p}$ and $h^{\pm}$ data, we will include all available SIA 
measurements from different experiments and with different 
center-of-mass energies. 
For the case of $\pi^{\pm}$, 
$K^{\pm}$ and $p/\bar{p}$, we use the data from the {\tt BELLE}, 
{\tt BABAR}, {\tt TASSO}, {\tt TPC}, {\tt TOPAZ}, {\tt ALEPH}, 
{\tt DELPHI}, {\tt OPAL} and {\tt SLD} Collaborations 
\cite{Leitgab:2013qh,Lees:2013rqd,Brandelik:1980iy,Althoff:1982dh,
Braunschweig:1990yd,Aihara:1988su,Braunschweig:1988hv,Buskulic:1994ft,
Buskulic:1995aw,Abreu:1998vq,Akers:1994ez,Ackerstaff:1998hz,Abe:2003iy}. 
These data are based on inclusive cross section measurements 
which contain all quark flavors, as well as flavor-tagged light 
($uds$)-, charm ($c$)-, and bottom ($b$)-quark samples. Note that 
constraints on heavy quark FFs is provided by the heavy flavor-tagged 
data.

For the case of unidentified light charged hadron 
$h^{\pm}$ data, we use the SIA measurements by the 
{\tt TASSO}, {\tt TPC}, {\tt ALEPH}, {\tt DELPHI}, {\tt OPAL} and 
{\tt SLD} Collaborations 
\cite{Buskulic:1995aw,Ackerstaff:1998hz,Abreu:1998vq,Aihara:1988su,
Abe:2003iy,Braunschweig:1990yd}. The SIA data included in our analysis 
are listed in Tables~\ref{tab:datasetsNLO} and \ref{tab:datasetsNNLO}. 
The second column of these tables contains the value of the 
center-of-mass energy for each experiment. The data cover 
center-of-mass energies between 10.52~GeV and 91.2~GeV.
The total number of data points included is 1492. 
This combines 392 data points for unidentified light charged hadrons 
$h^{\pm}$, 412 for pions, 369 for kaons and 319 for protons. 

The details of our fitting procedure will be discussed below, but 
we present already here, in the last four columns of  
Tables~\ref{tab:datasetsNLO} and \ref{tab:datasetsNNLO}, the values 
of $\chi^2$ per number of data points, $\chi^2/(N_\text{pts.})$, for 
each data set. The value of the total $\chi^2$ per number of degrees 
of freedom, $\chi^2/({\rm d.o.f.})$, is shown in the last line of 
these tables. It should be noted that the number of data points of 
each data set shown in the tables is subject to kinematic cuts. 
Actually, we remove data points at small- and large-$z$ in order 
to avoid regions where re-summation effects are sizeable.

We have examined a variety of kinematic cuts for different hadrons 
at small values of $z$. Since we include hadron mass effects in our 
analysis which could affect the small-$z$ region, we include more 
small-$z$ data points in our QCD fits than has been done in previous 
studies. Here we provide some details about the choice of the 
interval $\left[z_{\rm min}, z_{\rm max}\right]$ in which data 
points are included in our fit. In general, our choice for 
$z_{\rm min}$ and $z_{\rm max}$ varies with the center-of-mass 
energy. Choosing the same value of $z_{\rm min}$ = 0.02 for all 
experiments and for all center-of-mass energies leads to 
$\chi^2/{\rm d.o.f.}$ =1.415 and 1.228 for our NLO and NNLO analyses, 
respectively. Choosing the values of $z_{\rm min}$ = 0.075 instead 
of 0.02 leads to $\chi^2/{\rm d.o.f.}$=1.167 and 1.131 for the NLO and 
NNLO analyses, respectively. We found that it is indeed much better 
to include the data points with $z \ge 0.02$ for the center-of-mass 
energy of $\sqrt{s} = M_\texttt{Z}$, and $z \ge 0.075$ for $\sqrt{s} 
< M_\texttt{Z}$, where $M_\texttt{Z}$ is the mass of Z boson, for 
all different hadrons considered in the analysis. After imposing 
these kinematical cuts, we end up with a total of $N_\text{\rm pts.} 
= 1492$. As shown in Tables~\ref{tab:datasetsNLO} and 
\ref{tab:datasetsNNLO}, with these choices of kinematic cuts we 
find $\chi^2/{\rm d.o.f.} = 1.171$ for NLO and $\chi^2/{\rm d.o.f.} 
= 1.083$ for the NNLO fit, i.e.\ the NNLO fit shows in general a 
better fit quality.  

Compared with the most recent analysis by the {\tt NNFF1.0} 
collaboration~\cite{Bertone:2017tyb}, we use the same data sets for the 
identified light charged hadron production. However, our analysis 
is enriched with the additional unidentified light charged hadron 
production data sets. We agree with {\tt NNFF1.0} in the choice of 
$z_{\rm min}$: $z_{\rm min}$ = 0.02 for experiments at $\sqrt{s}= M_Z$, 
and  $z_{\rm min}$ = 0.075 for all other experiments. However, 
{\tt NNFF1.0} use only data up to $z_{\rm max}$ = 0.9 for all 
experiments.

%
\section{ The QCD framework for the {\tt SGKS20} FFs }
\label{sec:QCD_analysis}
%

In this section, we turn to present our theoretical framework to 
perform a simultaneous determination of charged pion, charged kaon 
and proton/antiproton FFs using the available SIA experimental 
data, together with data for unidentified light charged hadron 
production. 

It is, of course, impossible to determine a set of functions from 
a finite set of data points. One has to assume an ansatz which reduces 
the unknown functional dependence to a finite set of parameters. 
The particular choice is always a compromise between physical 
motivation and flexibility, and a certain amount of bias resulting 
from a too restrictive choice is unavoidable. 

In the present analysis, following Ref.~\cite{Soleymaninia:2019sjo}, 
we parameterize all the charged pion, charged kaon and proton/antiproton 
FFs at the input scale $\mu_0 = 5$ GeV, using the following 
functional form: 
\begin{eqnarray}
\label{input}
D^{H}_i(z, Q_{0})
= \frac{{\cal N}_i
z^{\alpha_i} (1 - z)^{\beta_{i}}
[1 + \gamma_{i}(1 - z)^{\delta_i}]}
{B[2 + \alpha_i, \beta_i + 1] +
\gamma_i B[2 + \alpha_i, \beta_i + \delta_i + 1]},
\nonumber 
\\
\end{eqnarray}
where $B[a,b]$ is the Euler Beta function, $H$ refers to the type of 
hadron, $H = \pi^{\pm}$, $K^{\pm}$ or $p/\bar{p}$, $i$ denotes the 
parton type, and ${\cal N}_i$ is the normalization constant for 
each flavor which is considered to be a fit parameter. 

Data provide information for inclusive and flavor tagged hadron production, 
i.e., we can expect that there is sufficient information to separate light 
flavor from charm- and bottom-quark initiated fragmentation. The separation 
of the gluon and the light-flavor FFs enters indirectly through the 
scale dependence. In particular, light flavors are separated by the fact 
that up- and down-quarks enter with scale-dependent coupling weights 
\cite{Bertone:2017tyb}. We therefore consider FFs for the flavor combinations 
$i = u^+$, $d^+$, $s^+$, $c^+$, $b^+$, and the gluon $g$, where $q^+ = 
q + \bar{q}$. SIA data allow us to consider only the sum of quark and 
anti-quark FFs since these data provide information on certain hadron 
species summed over the two charge states. 

For the $\pi^{\pm}$ FFs, we use isospin symmetry and relate 
$$
D^{\pi^\pm}_{u^+} 
= D^{\pi^\pm}_{d^+} \,. 
$$ 
For the proton/antiproton FFs, we parameterize $d^+$ and $s^+$ FFs, 
as described above in Eq.~(\ref{input}), but assume that the $u^+$ 
FF has the same shape as the $d^+$ FF, i.e.\ these two FFs are 
related by a $z$-independent normalization factor 
$\cal N$~\cite{deFlorian:2007ekg}, 
\begin{eqnarray}\label{input-proton}
D^{p/\bar{p}}_{u^+} =  {\cal N  } D^{p/\bar{p}}_{d^+}.
\end{eqnarray}
For the case of kaon FFs, one cannot assume that $u$ and $d$ FFs 
are related in the same way as for pions. The $d$ quark to kaon 
fragmentation is unfavored. We therefore allow all light-flavor kaon 
FFs to be different, 
$$ 
D^{K^\pm}_{u^+} 
\ne D^{K^\pm}_{d^+} 
\ne D^{K^\pm}_{s^+} \, .
$$
This parametrization with 6 independent kaon FFs provides us with 
additional flexibility and follows the choice of other studies 
\cite{Sato:2016wqj,Hirai:2007cx}, but differs from the one in 
Ref.~\cite{Bertone:2017tyb} for {\tt NNFF1.0} where only a 
5-component parametrisation for the kaon FFs was used. 

Data with unidentified light charged hadrons contain additional 
information which can provide further constraints on the determination 
of FFs. In our recent analysis of pion FFs \cite{Soleymaninia:2019sjo}, 
we could show that the inclusion of data for unidentified light charged 
hadrons affected the determination of pion FFs and has also led 
to a reduction of their uncertainties in some kinematic regions. 
We are therefore motivated to include unidentified hadrons also 
in the present analysis. 

By definition, unidentified light charged hadrons include $\pi^{\pm}$, 
$K^{\pm}$, $p/\bar{p}$, and an additional small \textit{residual} 
contribution from other light hadrons. Hence, the FFs of unidentified 
light charged hadrons is given by 
\begin{eqnarray}
\label{unidentified-def}
D_i^{h^\pm} =
D_i^{\pi^\pm} +
D_i^{K^\pm}
+ D_i^{p/\bar{p}} +
D_i^{{\it res}^\pm} \, . 
\end{eqnarray}
The residual light hadrons contribution is expected to be rather small. 
However, the most recent study in Ref.~\cite{Mohamaditabar:2018ffo} 
shows that the contribution from residual hadrons is significant for 
the case of $c$- and $b$-tagged cross sections. We consider a simple 
parametrization for the residual light hadron FFs $D^{{\it res}^\pm}$ 
as described in Ref.~\cite{Mohamaditabar:2018ffo}. It is given by 
\begin{eqnarray}
\label{eq:res}
D^{{\rm res}^\pm}_i(z, Q_0)
= {\cal N} _{i}
\frac{z^{\alpha_i}
(1 - z)^{\beta_{i}}} 
{B[2 + \alpha_{i},
\beta_{i} + 1]}\,,
\end{eqnarray}
where $i$ refers to  $u^+,\, d^+,\, s^+,\, c^+,\, b^+,\,$ and  $g$.
The normalization ${\cal N}_{i}$ of the FFs will be determined along 
with the other free parameters $\alpha_i$ and $\beta_i$ from the fit 
to the data. Since the analyzed SIA data are not sensitive to the 
separation of light quark flavors $(u,d,s)$, we assume an SU(3) 
flavor symmetric ansatz, 
$$ 
D^{{\rm res}^\pm}_{u^+}
= D^{{\rm res}^\pm}_{d^+}
= D^{{\rm res}^\pm}_{s^+} \, . 
$$
With these assumptions we have introduced 12 additional fit parameters 
for the residual light hadron FFs. 

The currently available SIA data do not fully constrain the entire 
$z$ dependence of quark and gluon FFs presented in Eqs.~\eqref{input} 
and \eqref{eq:res}.
Consequently, we are forced to make some further restrictions on the 
parameter space of the FFs. In particular, we found that the parameters 
$\gamma$ and $\delta$ are not well constrained by the SIA data. 
Therefore we consider them equal to zero for each flavor $i$ of the 
$K^{\pm}$ and $p/\bar{p}$ FFs, and also for the $s^+$, $c^+$ and $g$ 
FFs of $\pi^{\pm}$. To be more precise, just the $u^+$ and $b^+$ FFs 
of pions are considered to include five free parameters. In addition 
we found that the parameters $\alpha_{s^+}^{\pi^\pm}$, 
$\alpha_{s^+}^{K^\pm}$, $\beta_{c^+}^{p/\bar p}$ and  
$\beta_{b^+}^{p/\bar p}$ are not well constrained by SIA data and we 
have fixed them at their best values which were found in pre-fits. 
Finally, for the residual light hadron FFs, the parameters $\alpha$ 
and $\beta$ for the $u^+$, $d^+$, $s^+$, $c^+$ and the gluon FFs and 
$\alpha$ for the $b^+$ FF are only loosely determined by the fit and 
we fix them as well. 

We find that these restrictions of the shape parameters of FFs only 
marginally limit the freedom of the input functional form for the kaon 
and proton/antiproton FFs. In total, we have 54 free fit parameters 
which we include in the FFs uncertainty estimation.

Our results show that taking into account these residual contributions 
decreases the $\chi^2/{\rm d.o.f.}$ from 1.297 to 1.171 for the NLO 
analysis and from 1.261 to 1.083 for our NNLO analysis which in general 
indicates a better agreement of data and theory. 
We observe that the residual FFs obtained from the combined fit 
of the present work agree very well with the previous determination 
described in Ref.~\cite{Mohamaditabar:2018ffo} where the 
$D^{{\rm res}^\pm}_i$ have been found using {\tt NNFF1.0} FFs 
for the identified hadrons. 

As indicated, mass effects in pion, kaon and proton production 
are included in our QCD analysis. According to the definition of 
unidentified light charged hadrons in Eq.~\eqref{unidentified-def} 
and considering the fact that most of the contributions of light 
hadrons in unidentified light hadrons is relevant to the pion, kaon 
and proton, respectively, including their mass corrections is 
expected to improve the results, especially in the region of small 
$z$ and small $\sqrt{s}$. Hadron mass effects have been studied in 
Ref.~\cite{Nejad:2015fdh,Albino:2008fy} for $e^+e^-$ annihilation 
processes. We follow the strategy described in these references and 
incorporate hadron mass effects in single inclusive hadron production 
in SIA. For zero hadron mass, the scaling variable is expressed as 
$z = 2 E_H / \sqrt{s}$. A finite value of the hadron masses can be 
incorporated by a specific choice of the scaling variable. We define 
the light-cone scaling variable $\eta$ as 
\begin{eqnarray}
\label{eta}
\eta=
\frac{z}{2}
\left(1+\sqrt {1-
\frac{4m_H^2}
{sz^2}}\right),
\end{eqnarray}
where $m_H$ is the hadron mass. Accordingly, the differential cross 
section in the presence of hadron mass effects reads 
\begin{eqnarray}
\label{mass-cross}
\frac{d\sigma}{dz}=
\frac{1}
{1-\frac{m_H^2}{s\eta ^2}}
\sum_a \int_{\eta}^1
\frac{dx_a}{x_a}
\frac{\hat
{d\sigma_a}}{dx_a}
D_a^H\left(\frac{\eta}{x_a} , 
\mu\right) \,.
\end{eqnarray}
According to Eqs.~\eqref{eta} and \eqref{mass-cross}, the hadron mass 
corrections are most relevant in the small-$z$ and low-$\sqrt{s}$ 
regions. These formulas are applied for all three types of hadrons, 
i.e.\ pions, kaons and protons. The values of the hadron masses used 
in Eqs.~\eqref{eta} and \eqref{mass-cross} are considered to be 
$m_\pi= 0.140$~GeV, $m_K = 0.494$~GeV, and $m_p = 0.938$~GeV. 
We omit the hadron mass corrections for unidentified hadrons. 

We note that the effects of accounting for non-zero hadron masses 
in extracting the light hadron FFs have been explored recently also 
by {\tt NNFF1.0} for the case of pions, kaons, and protons 
FFs~\cite{Bertone:2017tyb}. It was observed that hadron-mass 
corrections can become significant in the kinematic region covered 
by the SIA data. Indeed, our present analysis confirms that hadron-mass 
corrections do improve the fit quality. Our detailed investigations 
show that ignoring these corrections in our QCD fits would lead to 
larger values of $\chi^2$. At NLO we find $\chi^2/{\rm d.o.f.} = 1.280$ 
and at NNLO $\chi^2/{\rm d.o.f.}$ = 1.241 if mass effects are omitted,  
while with mass effects included the corresponding values decrease 
to 1.171 and 1.083 for NLO and NNLO, respectively. 

In the present study, we use the publicly available {\tt APFEL} 
package~\cite{Bertone:2013vaa} for both evolving FFs and performing 
the numerical calculation of the SIA cross sections. Note that, using 
{\tt APFEL}, the related calculations can be performed up to NNLO 
accuracy in QCD. 
We should stress here that measurements of the longitudinal SIA 
cross-section ($d\sigma_L^{h^\pm}/dz$) are only available for the 
production of unidentified hadrons, $h^\pm$. However, one cannot 
analyze these data at NNLO as perturbative corrections to the 
coefficient functions are only available up the NLO accuracy in 
this case~\cite{Bertone:2018ecm}. Hence, we omit the data from the 
measurements of the longitudinal SIA cross-section. 
The effect of heavy quark masses are not 
taken into account in the present analysis and we use the zero mass 
variable flavor number scheme (ZM-VFNS) with five active flavors, 
including charm and bottom FFs. Moreover, the value of the strong 
coupling constant at the scale of the $Z$ boson mass is considered to 
be $\alpha_s (M_\texttt{Z}^2)= 0.118$~\cite{Tanabashi:2018oca}. For 
performing minimization and determination of fit parameters, we use 
the CERN program {\tt MINUIT}~\cite{James:1975dr}. The definition of 
$\chi^2$ is the same as the one we used in our previous 
works~\cite{Soleymaninia:2018uiv,Soleymaninia:2019sjo}, including the 
overall normalization errors of the experimental data sets. For 
calculating the uncertainties of the extracted FFs, we use the standard 
``Hessian'' approach~\cite{Martin:2009iq,Pumplin:2001ct} with 
$\Delta \chi^2 = 1$ (for further details, see 
Ref.~\cite{Soleymaninia:2018uiv}).
We will briefly describe our method of minimization and uncertainty 
estimation in the next section.  

\begin{table}
\begin{tabular}{lrrrr}
\hline
Parameter   & ~ $\pi^\pm$  & ~  $K^\pm$  & ~ $p/\bar{p}$ & ~ $res^\pm$  \\
				\hline \hline
				${\cal N}_{u^+}$&~ $0.9527$&~ $0.2531$&~ $0.8039$&~$0.0019$\\
				$\alpha_{u^+}$&~ $-0.7271$&~ $-0.8381$& ~$1.4098$&~$152.1475^*$\\
				$\beta_{u^+}$ &~ $1.6150$&~ $1.7252$&~ $5.3543$&~$15.0465^*$ \\
				$\gamma_{u^+}$ &~ $4.4861$&~ $0^*$&~ $0^*$&~$0^*$  \\
				$\delta_{u^+}$ &~ $3.6961$&~ $0^*$&~ $0^*$&~$0^*$  \\
				${\cal N}_{d^+}$&~ $0.9527$&~ $0.1551$&~ $0.0620$&~$0.0019$\\
				$\alpha_{d^+}$&~ $-0.7271$&~ $-0.4391$&~ $1.4098$&~$152.1475^*$\\
				$\beta_{d^+}$ &~ $1.6150$&~ $7.6257$&~ $5.3543$& ~$15.0465^*$\\
				$\gamma_{d^+}$ &~ $4.4861$&~ $0^*$&~ $0^*$&~ $0^*$  \\
				$\delta_{d^+}$ &~ $3.6961$&~ $0^*$&~ $0^*$&~ $0^*$  \\
				${\cal N}_{s^+}$&~ $0.7098$&~ $0.3125$& ~$0.0200$&~$0.0019$  \\
				$\alpha_{s^+}$&~ $0.0311^*$&~ $-0.5743^*$&~ $1.1364$&~$152.1475^*$  \\
				$\beta_{s^+}$ &~ $9.8675$&~ $2.0694$&~ $2.0407$&~$15.0465^*$  \\
				${\cal N}_{c^+}$&~ $0.7908$&~ $0.2770$&~ $0.0198$&~$0.030$  \\
				$\alpha_{c^+}$&~ $-0.7437$&~ $-0.3101$&~ $10.8627$&~$5.6831^*$  \\
				$\beta_{c^+}$ &~ $5.7138$&~ $4.9055$&~ $52.8237^*$&~$11.7035^*$ \\
				${\cal N}_{b^+}$&~ $0.7499$&~ $0.2175$& ~$0.0049$&~$0.1082$\\
				$\alpha_{b^+}$&~ $-0.2896$&~ $0.2811$&~ $3.8762$&~$1.6225^*$  \\
				$\beta_{b^+}$ &~ $5.2067$&~ $12.2417$&~ $159.332^*$&~$6.5464$  \\
				$\gamma_{b^+}$ &~ $9.6277$&~ $0^*$&~ $0^*$&~ $0^*$  \\
				$\delta_{b^+}$ &~ $8.8143$&~ $0^*$&~ $0^*$&~ $0^*$ \\
				${\cal N}_{g}$&~ $0.4801$&~ $0.1018$& ~$0.1910$&~$0.0270$ \\
				$\alpha_{g}$&~ $1.5868$&~ $9.5790$&~ $2.3699$&~$20.4675^*$  \\
				$\beta_{g}$ &~ $29.8298$&~ $7.5076$&~ $7.6487$&~$13.8349^*$  \\ 	
				\hline 		 	\hline 	
\end{tabular}
\caption{ 
  Best-fit parameters for the fragmentation of partons into $\pi^\pm$,  
  $K^\pm$, $p/\bar{p}$ and residual light charged hadrons ($res ^\pm$) 
  obtained through a simultaneous analysis at 
  NLO accuracy within the framework described in 
  Sec~\ref{sec:QCD_analysis}. The starting scale has been taken to be 
  $\mu_0=5$ GeV for all parton species. Parameters marked with an 
  asterisk are fixed input quantities.
  }
\label{table:parsNLO}
\end{table}

\begin{table}
\begin{tabular}{lrrrr}
\hline
Parameter   & ~ $\pi^\pm$  & ~  $K^\pm$  & ~ $p/\bar{p}$ & ~ $res^\pm$ \\
				\hline \hline
				${\cal N}_{u^+}$&~ $0.9243$&~ $0.2409$&~ $0.7188$&~$0.0019$\\
				$\alpha_{u^+}$&~ $-0.8411$&~ $-0.7248$& ~$0.6275$&~$144.9869^*$\\
				$\beta_{u^+}$ &~ $1.7556$&~ $2.0895$&~ $4.8433$ &~$16.5308^*$\\
				$\gamma_{u^+}$ &~ $3.2186$&~ $0^*$&~ $0^*$ &~ $0^*$ \\
				$\delta_{u^+}$ &~ $4.3105$&~ $0^*$&~ $0^*$ &~ $0^*$ \\
				${\cal N}_{d^+}$&~ $0.9243$&~ $0.2486$&~ $0.0860$&~ $0.0019$\\
				$\alpha_{d^+}$&~ $-0.8411$&~ $-0.6878$&~ $0.6275$&~ $144.9869^*$\\
				$\beta_{d^+}$ &~ $1.7556$&~ $5.6757$&~ $4.8433$&~ $16.5308^*$ \\
				$\gamma_{d^+}$ &~ $3.2186$&~ $0^*$&~ $0^*$&~ $0^*$  \\
				$\delta_{d^+}$ &~ $4.3105$&~ $0^*$&~ $0^*$&~ $0^*$  \\
				${\cal N}_{s^+}$&~ $0.8006$&~ $0.2614$& ~$0.0162$&~ $0.0019$  \\
				$\alpha_{s^+}$&~ $-0.1781^*$&~ $-0.6810^*$&~ $0.6308$&~ $144.9869^*$ \\
				$\beta_{s^+}$ &~ $8.1331$&~ $1.6131$&~ $1.8532$ &~ $16.5308^*$ \\
				${\cal N}_{c^+}$&~ $0.8070$&~ $0.2836$&~ $0.0369$&~ $0.0291$  \\
				$\alpha_{c^+}$&~ $-0.8247$&~ $-0.4406$&~ $3.6331$&~ $9.8796^*$  \\
				$\beta_{c^+}$ &~ $5.6455$&~ $4.7087$&~ $25.0310^*$&~ $19.1145^*$ \\
				${\cal N}_{b^+}$&~ $0.7686$&~ $0.2279$& ~$0.0058$&~ $0.1246$\\
				$\alpha_{b^+}$&~ $-0.3955$&~ $0.1040$&~ $3.3027$&~ $0.5507^*$  \\
				$\beta_{b^+}$ &~ $4.9983$&~ $11.4295$&~ $166.0012^*$&~ $5.6387$  \\
				$\gamma_{b^+}$ &~ $9.2937$&~ $0^*$&~ $0^*$&~ $0^*$  \\
				$\delta_{b^+}$ &~ $8.7525$&~ $0^*$&~ $0^*$&~ $0^*$ \\
				${\cal N}_{g}$&~ $0.4669$&~ $0.0884$& ~$0.1986$&~ $0.0115$ \\
				$\alpha_{g}$&~ $0.7742$&~ $12.1509$&~ $-0.1871$ &~ $24.6488^*$ \\
				$\beta_{g}$ &~ $24.7398$&~ $8.6869$&~ $3.7138$&~ $11.2409^*$  \\ 	
				\hline 	  	\hline 		
\end{tabular}
\caption{ 
  Same as Table.~\ref{table:parsNLO} but at NNLO accuracy. 
  }
\label{table:parsNNLO}
\end{table}

The best values of the fit parameters for charged pion, charged kaon, 
proton/antiproton and residual light hadrons FFs determined at the 
initial scale $\mu_0=5$ GeV are listed in Tables~\ref{table:parsNLO} 
and \ref{table:parsNNLO} at NLO and NNLO accuracy, respectively. 
Note that the parameters labeled with an asterisk ($^*$) are either 
fixed input quantities, or have been determined in a pre-fit and 
are kept fixed in the final fit to determine the other fit 
parameters and their uncertainty ranges.

\begin{figure*}[htb]
\resizebox{0.95\textwidth}{!}{\includegraphics{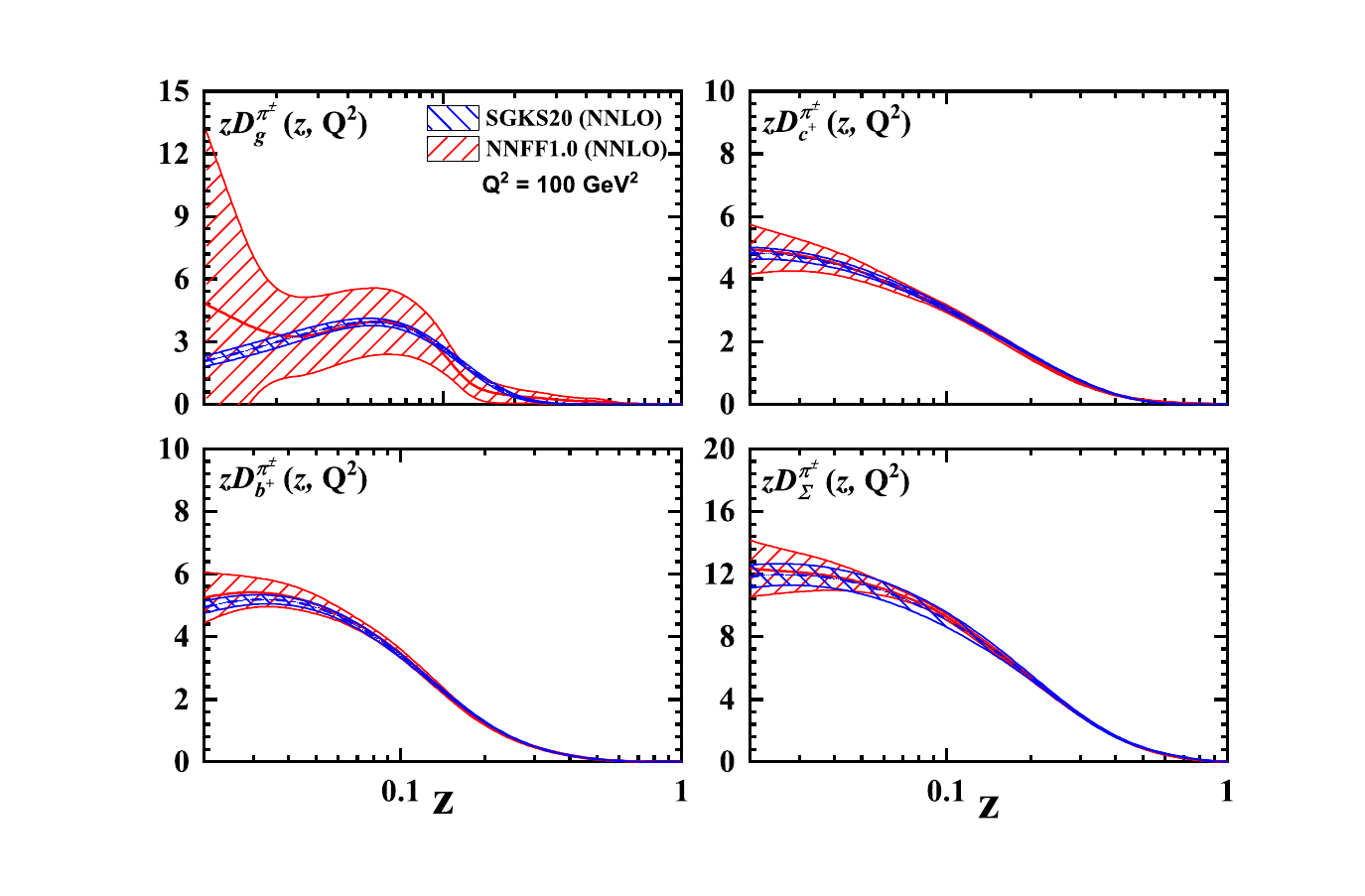}	}	
\begin{center}
\caption{ \small 
Comparison of {\tt SGKS20} NNLO charged pion FFs, 
$zD^{\pi^\pm}_i (z, Q^2)$ ($i = g, c, b, \Sigma$) 
together with their 1-$\sigma$ 
uncertainty bands at Q$^2 = 100$ GeV$^2$ with the 
results from the 
{\tt NNFF1.0} Collaboration. 
}
\label{fig:compare-pion-with-NNFF}
\end{center}
\end{figure*}
\begin{figure*}[htb]
\resizebox{0.95\textwidth}{!}{\includegraphics{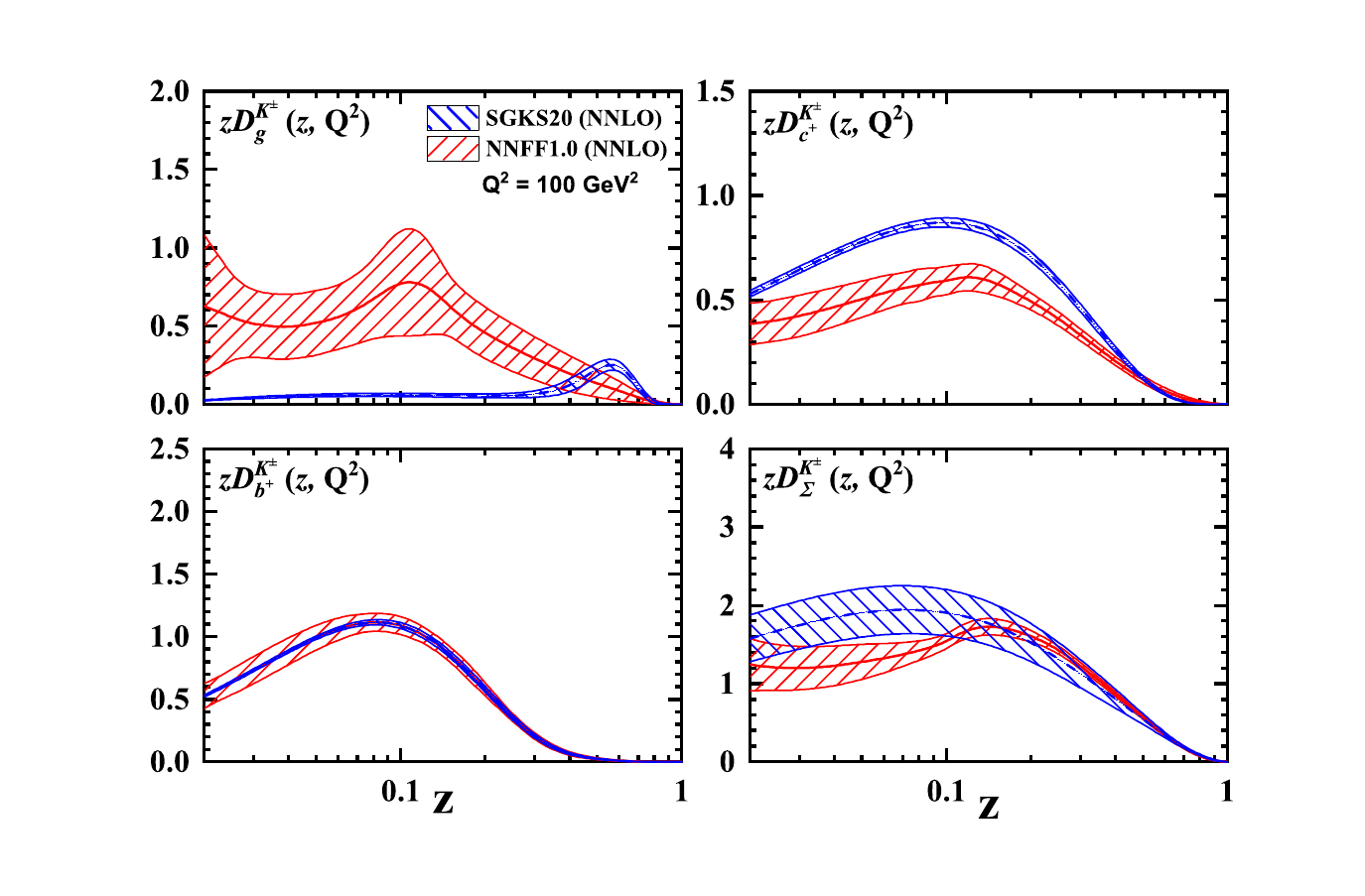}  }		
\begin{center}
\caption{ \small 
Same as Fig.~\ref{fig:compare-pion-with-NNFF} but for the 
charged kaon $zD^{K^\pm}_i (z, Q^2)$ FFs. 
}
\label{fig:compare-kaon-with-NNFF}
\end{center}
\end{figure*}
\begin{figure*}[htb]
\resizebox{0.95\textwidth}{!}{\includegraphics{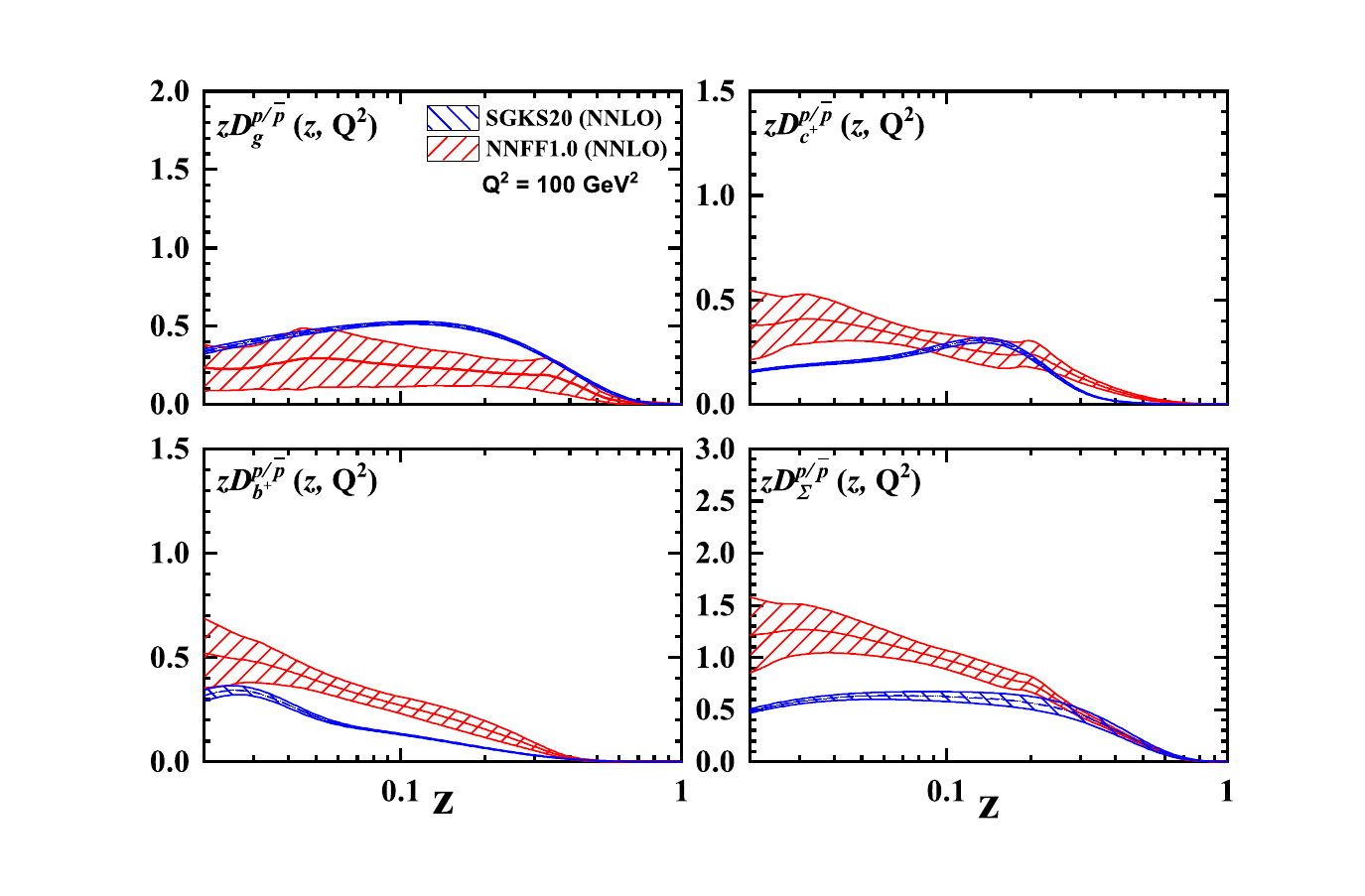} 	}	
\begin{center}
\caption{ \small 
Same as Fig.~\ref{fig:compare-pion-with-NNFF} but for the 
proton and antiproton FFs, $zD^{p/\bar{p}}_i (z, Q^2)$. 
}
\label{fig:compare-proton-with-NNFF}
\end{center}
\end{figure*}

\begin{figure*}[htb]
\vspace{0.50cm}
\resizebox{0.90\textwidth}{!}{\includegraphics{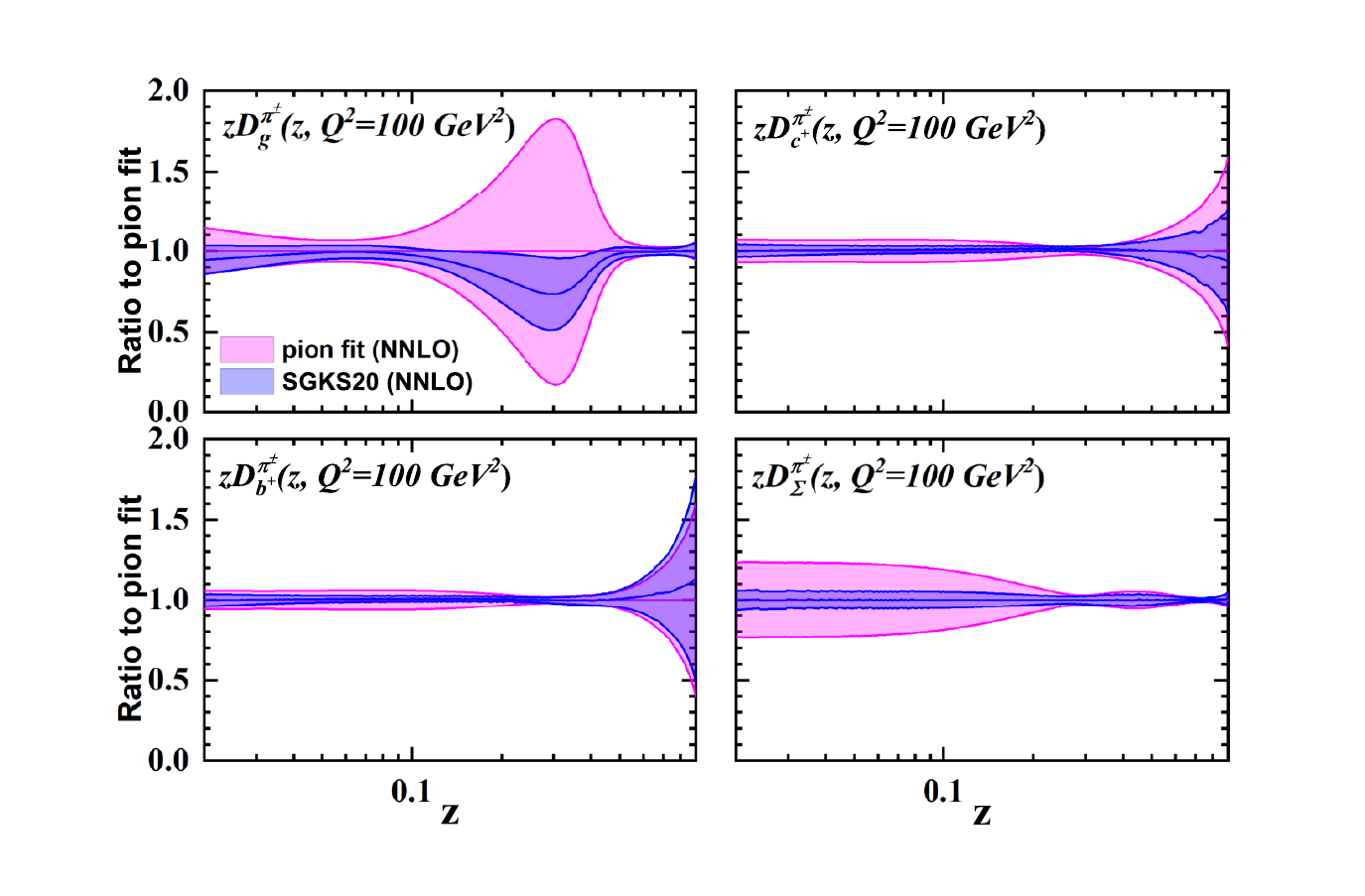}}
\begin{center}
\caption{{\small  
Comparison of {\tt SGKS20} NNLO charged pion FFs, 
$zD^{\pi^\pm}_i (z, Q^2 = 100~\text{GeV}^2)$ ($i = g, c, b, 
\Sigma$) presented in this study with results extracted from 
a fit without including unidentified light charged hadron data.   
}
\label{fig:pion-FFs-ratio}}
\end{center}
\end{figure*}
\begin{figure*}[htb]
\vspace{0.50cm}
\resizebox{0.90\textwidth}{!}{\includegraphics{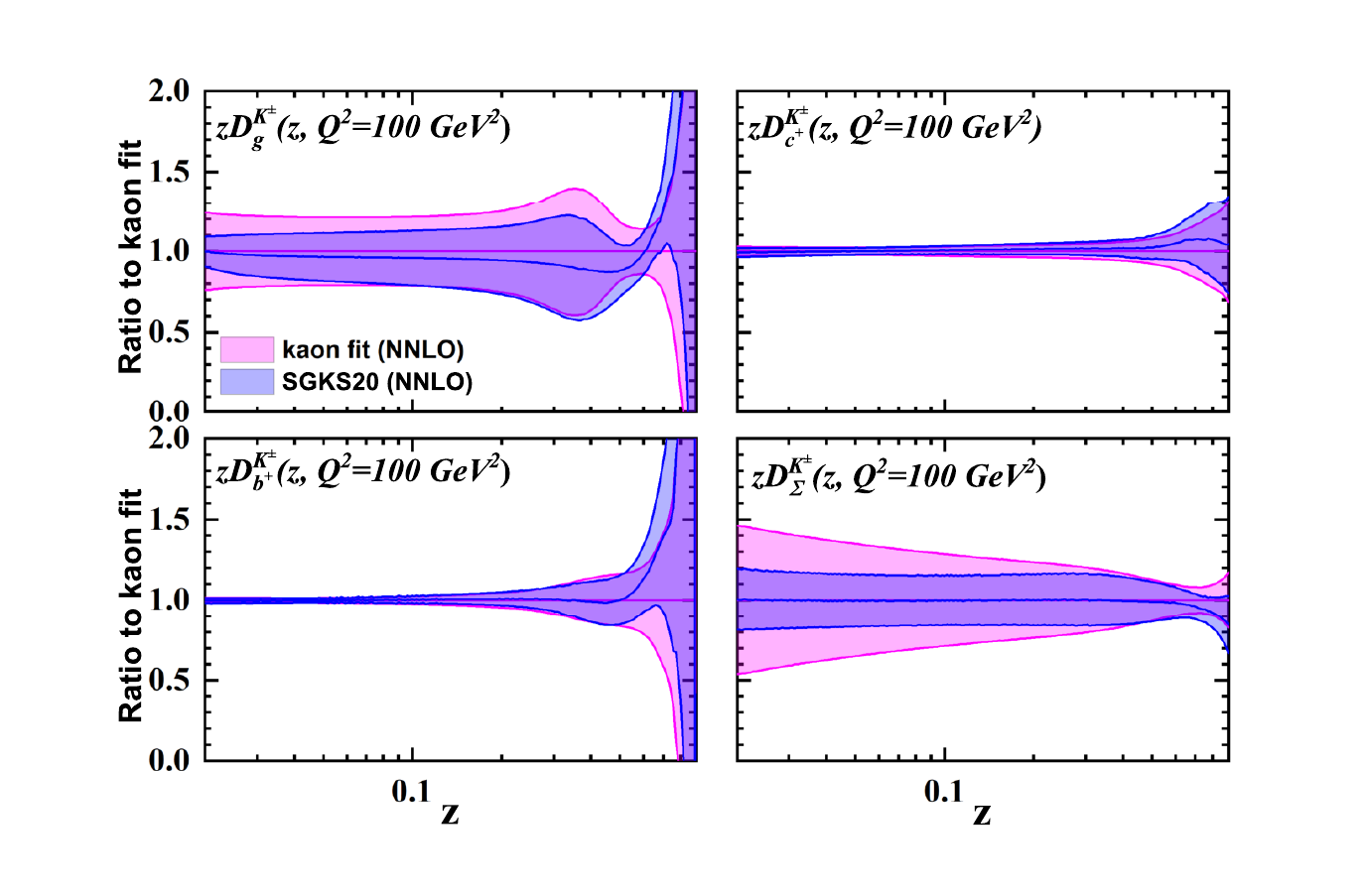}}
\begin{center}
\caption{{\small  
Same as Fig.~\ref{fig:pion-FFs-ratio} but for the 
charged kaon $zD^{K^\pm}_i (z, Q^2)$ FFs. 
}
\label{fig:kaon-FFs-ratio}}
\end{center}
\end{figure*}
\begin{figure*}[htb]
\vspace{0.50cm}
\resizebox{0.90\textwidth}{!}{\includegraphics{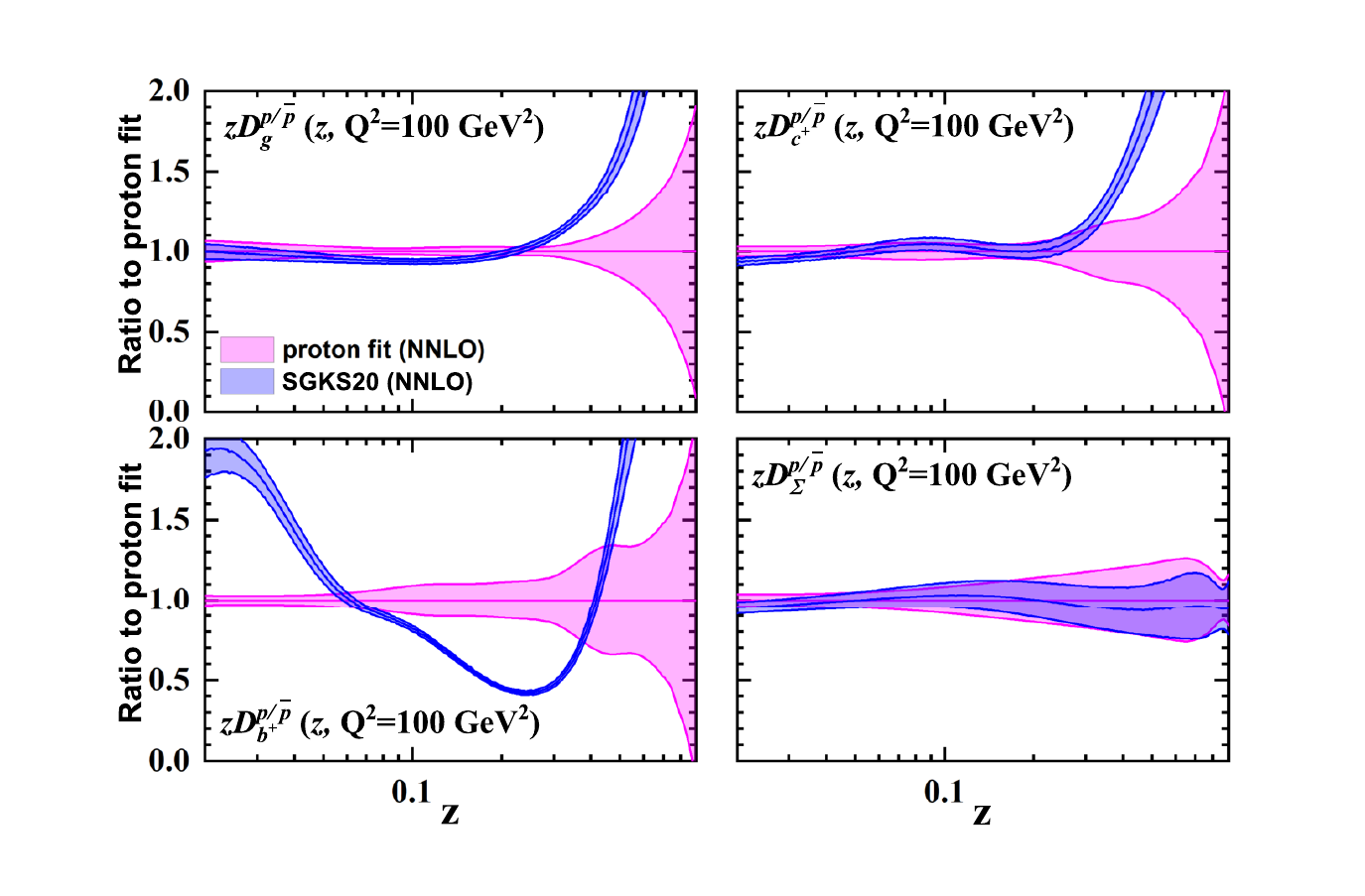}}
\begin{center}
\caption{{\small  
Same as Fig.~\ref{fig:pion-FFs-ratio} but for the 
proton and antiproton FFs, $zD^{p/\bar{p}}_i (z, Q^2)$. 
}
\label{fig:proton-FFs-ratio}}
\end{center}
\end{figure*}
\begin{figure*}[htb]
\vspace{0.50cm}
\resizebox{0.90\textwidth}{!}{\includegraphics{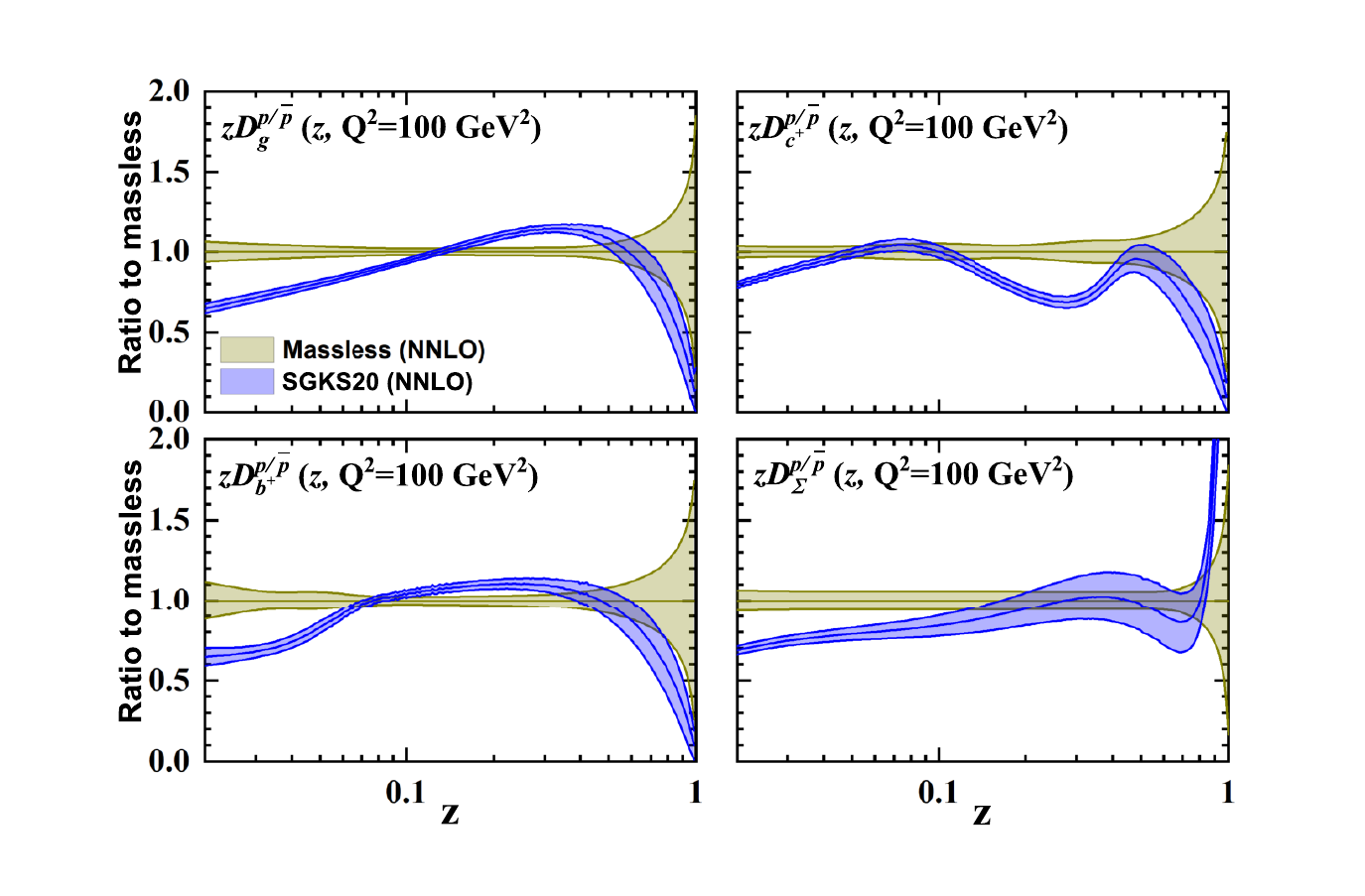}}
\begin{center}
\caption{{\small  
Comparison of {\tt SGKS20} NNLO proton and antiproton FFs 
presented in this study with results extracted from a QCD 
analysis without including hadron mass corrections.   
}
\label{fig:proton-FFs-ratio-Massless}}
\end{center}
\end{figure*}

\section{ $\chi^2$ minimization and uncertainty estimation }
\label{sec:minimizations}

The best values of the independent fit parameters defined in 
Eq.~\eqref{input} need to be determined from SIA data
by performing a minimization procedure using an effective $\chi^{2}$ function.
This function quantifies the goodness of fit 
to the SIA data for a given set of fit parameters, \{$p_{i}$\}.
The simplest method to calculate the total $\chi^{2}(\{p_{i}\})$ for a
 set of independent fit parameters $\{p_{i}\}$ is given by,
\begin{eqnarray}
\label{eq:chi2-1}
\chi^{2} (\{p_{i}\}) = 
\sum_{i}^
{n^{\text{data}}}
\frac{({\cal O}^
{\text {data}}_{i}
-{\cal T}^
{\rm{theory}}_i
(\{p_{i}\}))^{2}} 
{(\sigma^{\text{data}}
_{i})^{2}}\,,
\end{eqnarray}
where ${\cal O}^{\text{data}}_{i}$ refer to the experimental observables, 
and ${\cal T}^{\text{theory}}_{i}$ indicate the corresponding theoretical 
values at a given $z$ and $\mu^{2}$. In this study, the experimental 
errors, $\sigma^{\text{data}}_{i}$, in the above equation are calculated 
from the statistical and systematical errors added in quadrature. 
However, the analyses available in the  
literature~\cite{Martin:2009iq,Kovarik:2010uv,Kovarik:2015cma,Schienbein:2009kk}  
have shown that such a simple $\chi^{2}(\{p_{i}\})$ definition needs 
to be modified to account for correlations in the experimental 
uncertainties. In particular, most of the SIA data come with an overall 
normalization uncertainty which is fully correlated within one data 
set, but uncorrelated between different data sets. Therefore we split 
the global $\chi^{2}_{\text{global}}(\{p_{i}\})$ into contributions 
from individual data sub-sets, 
\begin{equation}
\label{eq:chi2-2}
\chi_{\text{global}}^{2}
(\{p_{i}\}) =
\sum_{n=1}^{{n}^
{\text{exp}}} \,
{\cal W}_{n} \, 
\chi_n^2 (\{p_{i}\})\,,
\end{equation}
where ${n}^{\text{exp}}$ is the number of individual experimental data 
sets and ${\cal W}_{n}$ indicates a weight factor for the $n^{\text{th}}$ 
experiment. Then, $\chi_{n}^{2} (\{p_{i}\})$ defined in 
Eq.~\eqref{eq:chi2-1} needs to be corrected as 
\begin{eqnarray}
\label{eq:chi2-3}
\chi_{n}^{2}
(\{p_{i}\})&=&
\left(\frac{1 - 
{\cal N}_{n}}
{\Delta{\cal N}_{n}}
\right)^{2} +
\nonumber \\
&&
\hspace{-0.5cm}
\sum_{k=1}^
{N_{n}^{\text{data}}}
\left(\frac{({\cal N}_{n}  \,
{\cal O}_{k}^
{\text{data}}-
{\cal T}_{k}^
{\text{theory}}
(\{p_{i}\})}
{{\cal N}_{n} \,
\delta{D}_{k}^
{\text{data}}}
\right)^{2}\,,
\end{eqnarray}
in which $i$ runs over all data points and ${N}^{\text{data}}_{n}$ 
corresponds to the number of data points in each data set. In order to 
determine the best fit parameters of the {\tt SGKS20} light charged 
hadrons FFs, we minimize the $\chi^{2}_{\text{global}}(\{p_{i}\})$ 
function with respect to the fit parameters $\{p_{i}\}$ presented in 
Eq.~\eqref{input}. The normalization factors $\Delta{\cal N}_{n}$ need 
to be fitted along with the independent fit parameters $(\{p_{i}\})$ 
and then can be kept fixed. The default value of the weight factors for 
each experimental data set is considered to be equal to 
1~\cite{Stump:2001gu,Blumlein:2006be}.

In the following, we briefly discuss our method to estimate the 
uncertainties of the {\tt SGKS20} light charged hadrons FFs. 
Three different approaches are available in the literature to estimate 
the uncertainty. They are based on Lagrange multipliers or Monte-Carlo 
sampling, but the most commonly used method is the so-called `Hessian' 
approach~\cite{Martin:2009iq}. Following the notation adopted in 
Refs.~\cite{Pumplin:2000vx,Pumplin:2001ct}, our uncertainty estimation 
is done using the standard 'Hessian' approach. In this method, the 
uncertainty for a fragmentation function, $\Delta D(z)$, can be obtained 
from linear error propagation. It is given by 
\begin{eqnarray}
\label{eq:Hessian}
&&
[\Delta D (z)]^{2}
= \Delta 
\chi^{2}_{\text{global}}
\times \, \nonumber \\
&& 
\left[ \sum_{i}^{n} 
\left(\frac{\partial
D(z, {\hat p})}
{\partial p_{i}}\right)^{2} \,
C_{i i}  + 
\sum_{i \neq j}^{n} 
\left( \frac{\partial 
D(z, {\hat p})}
{\partial p_{i}}
 \frac{\partial  
D(z, {\hat p})}
{\partial p_{j}} \right) \,  
C_{i j}  \right], 
\nonumber \\
\end{eqnarray}
where $p_{i}$ (with $i$ = 1, 2, ..., $n$) denotes the independent 
free parameters for each FF, $n$ refers to the total number of 
optimized parameters, and ${\hat p}_{i}$ comprises the numerical 
values of the optimized parameters. $C_{i, j} \equiv H_{i, j}^{-1}$ 
are the elements of the covariance matrix determined in this analysis 
at the input scale. In order to estimate the uncertainties of the 
{\tt SGKS20} light charged hadrons FFs, we follow the standard 
parameter-fitting criterion by considering contours of $T = 
\Delta \chi^{2}_{\text{global}} = 1$ defining the 68\% (1-$\sigma$) 
confidence level (CL). For minimization and the determination of both 
fit parameters and elements of the covariance matrix we use the publicly 
available CERN program {\tt MINUIT}~\cite{James:1975dr}.

%
\section{Results of the {\tt SGKS20} FF analysis } 
\label{sec:Results}
%

The following part of this article describes in greater detail the 
results of the {\tt SGKS20} FFs analysis. We focus on the inclusion 
of higher-order QCD corrections in our NNLO results. We also compare 
our best fit pion, kaon and proton/antiproton FFs with their 
counterparts from the {\tt NNFF1.0} analysis~\cite{Bertone:2017tyb}.

In Tables~\ref{table:parsNLO} and \ref{table:parsNNLO} we present 
the best fit parameters for the fragmentation functions of partons 
into $\pi^\pm$, $K^\pm$, $p/\bar{p}$ and the residual FFs at NLO and 
NNLO accuracy, respectively.

The NNLO charged hadron FFs, $zD^{H^\pm}_i (z, Q^2)$, for singlet 
($D^{H^\pm}_\Sigma = \sum _{q} (D^{H^\pm}_q+D^{H^\pm}_{\bar{q}})$,~ 
$q = u$, $d$, $s$), heavy-quark- and gluon-to-hadron fragmentation 
obtained from the combined fit are illustrated in 
Figs.~\ref{fig:compare-pion-with-NNFF}, 
\ref{fig:compare-kaon-with-NNFF} and \ref{fig:compare-proton-with-NNFF} 
together with their 1-$\sigma$ uncertainty bands for charged pions, 
charged kaons and protons/antiprotons, respectively.
The NNLO results from the most recent determination available in the 
literature, namely the {\tt NNFF1.0} FFs, are also shown for 
comparison. The results at Q$^2 = 100$ GeV$^2$ reveal the following 
findings. A noticeable feature of the distributions shown in 
Fig.~\ref{fig:compare-pion-with-NNFF} is the remarkable agreement  
between our $zD^{\pi^\pm} (z, Q^2)$ FFs for heavy and singlet quarks 
with the corresponding results from {\tt NNFF1.0}.
Fig.~\ref{fig:compare-pion-with-NNFF} shows a small difference for the 
gluon density, especially at small values of $z$. A further remarkable 
aspect of the comparison in this figure is related to the size of the 
uncertainties. For all cases, the {\tt SGKS20} 1-$\sigma$ error bands 
are smaller than those of the corresponding {\tt NNFF1.0} FFs.

Our charged kaon $zD^{K^\pm} (z, Q^2)$ FFs at NNLO accuracy are shown 
in Fig.~\ref{fig:compare-kaon-with-NNFF} in comparison with the 
{\tt NNFF1.0} FFs. Concerning the shapes of the kaon FFs, a number 
of interesting differences between  the {\tt SGKS20} and {\tt NNFF1.0} 
FFs can be seen. The differences in shape among the two FF sets are 
more marked than in the case of the charged pion FF. Moderate differences 
are observed for the central value of the singlet FF at smaller values 
of $z$, especially at $z<0.1$, and for the uncertainty band of the 
bottom FF below $z<0.05$. A more noticeable difference in shape is 
observed for the gluon and charm FFs for which the {\tt SGKS20} 
results are more suppressed and enhanced, respectively, at 
$z<0.4$, than  the gluon and charm FFs from {\tt NNFF1.0}.

Let us now discuss the {\tt SGKS20} protons and antiprotons 
$zD^{p/\bar{p}} (z, Q^2)$ FFs. A fair agreement is observed only in 
the case of the heavy-quark and singlet-quark FFs at large values of 
$z$. These FFs are more suppressed at medium to small $z$ values,  
compared with the corresponding FFs from {\tt NNFF1.0}. For 
$zD^{p/\bar{p}}_g$, big differences can be seen both in the magnitude 
and the error band of the FFs in the whole range of $z$. Overall, the 
error bands for all heavy quark, singlet and gluon FFs for all light 
hadrons are dramatically reduced, except for the singlet FF of the 
kaon at medium to large $z$.

There are a number of similarities and differences between 
the {\tt SGKS20} and {\tt NNFF1.0} analyses. The QCD approach used 
in this study is similar to the one used by {\tt NNFF1.0}. In both 
cases, NNLO QCD and hadron-mass corrections are taken into 
account. Also, the kinematic cuts imposed on data points in the 
small $z$ region are the same in both analyses. The origin of the 
differences between the {\tt SGKS20} and {\tt NNFF1.0} FFs is likely 
to be due to the following reasons.

First, the {\tt NNFF1.0} approach is based on neural networks without 
fixing a priori a specific parametrization. This allows one to obtain 
much more flexibility in the description of FFs. In contrast, 
{\tt SGKS20} uses the standard Hessian method with the choice of 
tolerance $\Delta \chi^2 = 1$ at 68\% confidence level. It can, 
therefore, be expected that the uncertainties of the {\tt NNFF1.0} 
FFs are larger than those of {\tt SGKS20}. This is indeed the case, 
as seen in the figures. Second, the origin of differences in the 
shape and error bands for the {\tt SGKS20} FFs is also due to the 
fact that we include more data in the analysis: data for unidentified 
light charged hadrons are taken into account along with identified 
$\pi^\pm$, $K^\pm$ and $p/\bar{p}$ production data, simultaneously in 
one fit. 

In the following, we present a systematic study in order to 
investigate in more detail the origin of differences between our 
results and {\tt NNFF1.0}. We will quantify the additional 
constraints due to the inclusion of unidentified light charged 
hadron data. To do so, we have extracted FFs from QCD analyses in 
which we excluded unidentified light charged hadron data from the 
fit, i.e.\ for each of the pion, kaon, and proton FFs we performed 
separate fits using only data for the respective hadron species. 
We present the results in terms of ratios where all FFs are 
normalized to their central values obtained in the separate-hadron 
fits. In Figs.~\ref{fig:pion-FFs-ratio}, \ref{fig:kaon-FFs-ratio} 
and \ref{fig:proton-FFs-ratio} the FFs for $\pi^\pm$, $K^\pm$ and 
$p/\bar{p}$ are shown at the reference scale $Q^2 = 100$~GeV$^2$. 

As one can see, the inclusion of unidentified light charged hadron 
data affect both the shape and the uncertainties of FFs. For all 
cases, by adding unidentified light charged hadron data, smaller 
uncertainties are obtained. For the case of pion FFs in 
Fig.~\ref{fig:pion-FFs-ratio}, the reduction of the uncertainty 
bands is clearly visible. The inclusion of unidentified light 
charged hadron data also affects the shape of the gluon FF of pions. 
These findings are in good agreement with our previous 
study~\cite{Soleymaninia:2019sjo} where we had examined the effect 
of such data on the determination of pion FFs.

Similarly, results for the case of charged kaon FFs are presented 
in Fig.~\ref{fig:kaon-FFs-ratio}. Here again, one can conclude that 
the inclusion of unidentified light charged hadron data leads to a 
reduction of the uncertainties in the case of $D^{K^\pm}_g$ and 
$D^{K^\pm}_\Sigma$, although by a smaller factor than in the case 
of the pion FFs.

In the case of proton and antiproton FFs, $D^{p/\bar{p}}_i$, we 
find particularly significant changes of the shape between the
``proton fit'' and the combined {\tt SGKS20} analysis, except for 
the case of $D^{p/\bar{p}}_\Sigma$. The gluon and the heavy-quark 
FFs are strongly affected at large $z$, while the $b^+$ FF changes 
over the whole range of $z$. As can be seen in 
Fig.~\ref{fig:proton-FFs-ratio}, adding unidentified light charged 
hadron data also leads to a large reduction of the uncertainty 
bands, again with the exception of the $\Sigma$ FF. 

Now we also want to discuss in detail the effect arising from the 
inclusion of hadron mass corrections on the shape and uncertainties 
of FFs. In Fig.~\ref{fig:proton-FFs-ratio-Massless}, we compare the 
{\tt SGKS20} NNLO proton and antiproton FFs presented in this study 
with the corresponding results that have been extracted from the QCD 
analysis in which we exclude hadron mass corrections. Since the mass 
of the proton is larger than those of the pion and kaon, the effect 
of hadron mass corrections is expected to be most important for the 
proton FFs. Hence, we present our results for this case only.

Concerning the shapes of the $p/\bar{p}$ FFs, a number of interesting
differences between the two results can be seen from the comparisons 
in Fig.~\ref{fig:proton-FFs-ratio-Massless}. The inclusion of proton 
mass corrections affects both the shape and the uncertainty of 
$p/\bar{p}$ FFs. In particular, the low $z$ region is strongly 
affected in all cases. There is also a slight reduction of the 
uncertainties at low $z$, but this is not particularly strong. 

As a short summary, our systematic study has shown that there are 
significant changes of the FF fit results due to the inclusion of 
unidentified light charged hadrons. This can explain part of the 
differences between the {\tt SGKS20} and {\tt NNFF1.0} fits. A 
detailed comparison of the results shown in 
Figs.~\ref{fig:compare-pion-with-NNFF}, 
\ref{fig:compare-kaon-with-NNFF}, \ref{fig:compare-proton-with-NNFF} 
with those in Figs.~\ref{fig:pion-FFs-ratio}, 
\ref{fig:kaon-FFs-ratio}, \ref{fig:proton-FFs-ratio} allows us to 
conclude that a large part of the differences of the width of the 
uncertainty bands is, however, more likely due to the different 
fit methodology, i.e.\ due to the fact that we use the Hessian 
method with a $\chi^2$-based definition of a confidence interval.

Considering the fit quality upon inclusion of higher-order QCD 
corrections, one can conclude from Tables~\ref{tab:datasetsNLO} and 
\ref{tab:datasetsNNLO} that the NNLO corrections slightly improve the 
overall fit quality for almost all SIA data. As one can see from 
these tables, the  $\chi^2/({\rm d.o.f.})$ values at NNLO accuracy 
are lower than at NLO. Moreover, the fit quality suggests that the 
inclusion of \textit{residual} light-hadron contributions as well 
as unidentified light charged hadron data in our identified 
$zD^{H^\pm} \, (H^\pm = \pi^\pm, K^\pm, p/\bar{p})$ analysis leads to 
an improved agreement between theory and data.

\begin{figure*}[htb]
\vspace{0.50cm}
\resizebox{0.480\textwidth}{!}{\includegraphics{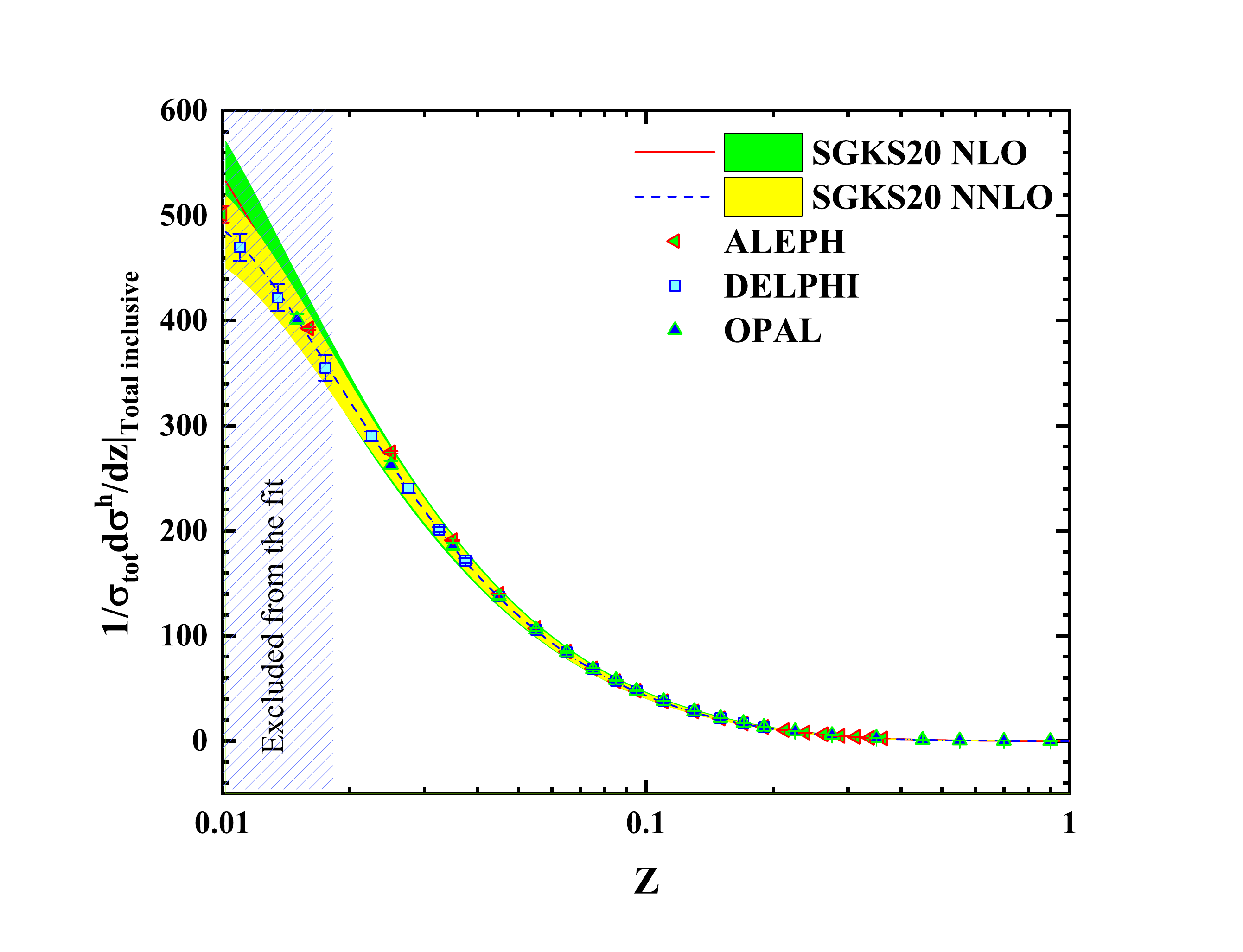}} 	
\resizebox{0.480\textwidth}{!}{\includegraphics{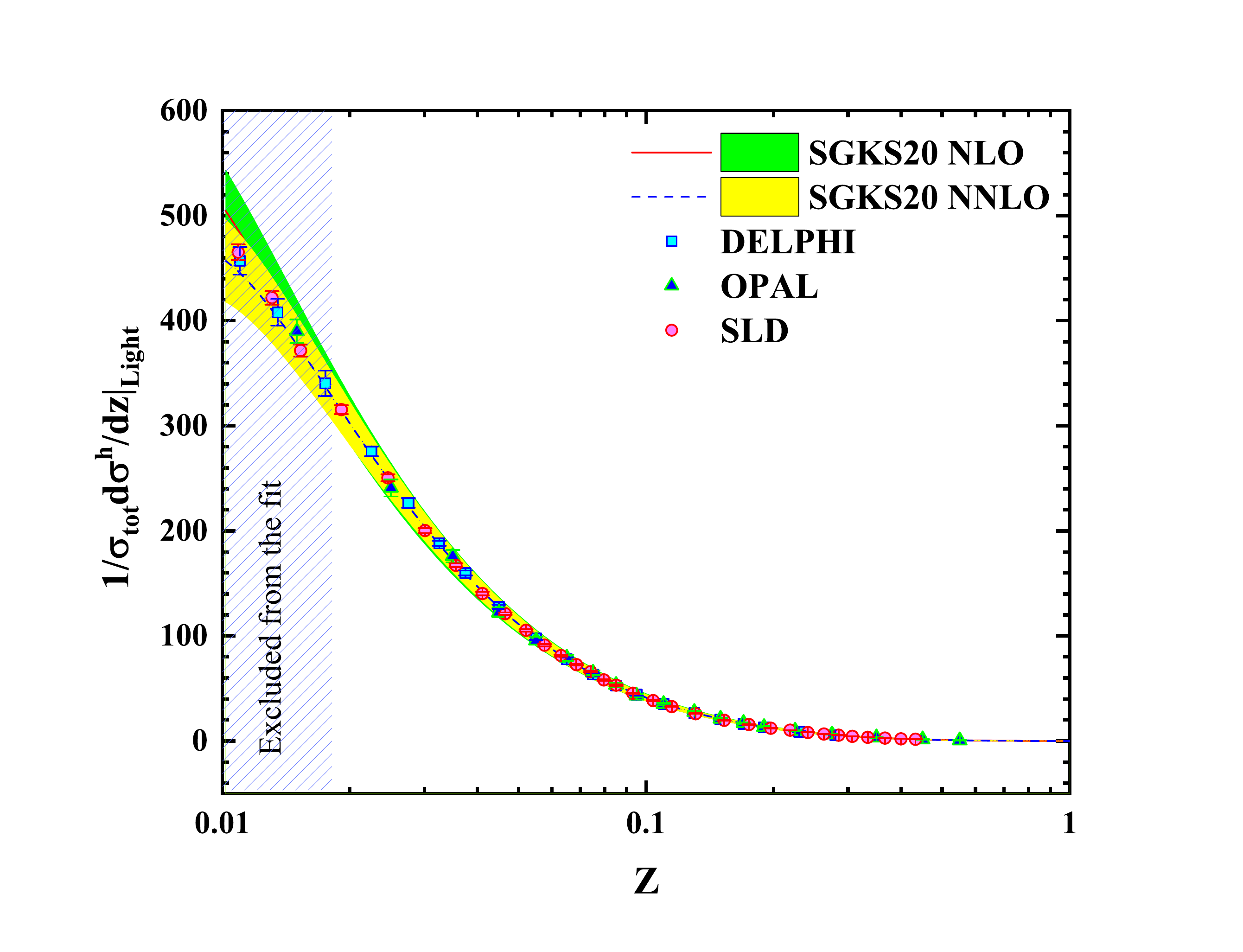}}  	
\resizebox{0.480\textwidth}{!}{\includegraphics{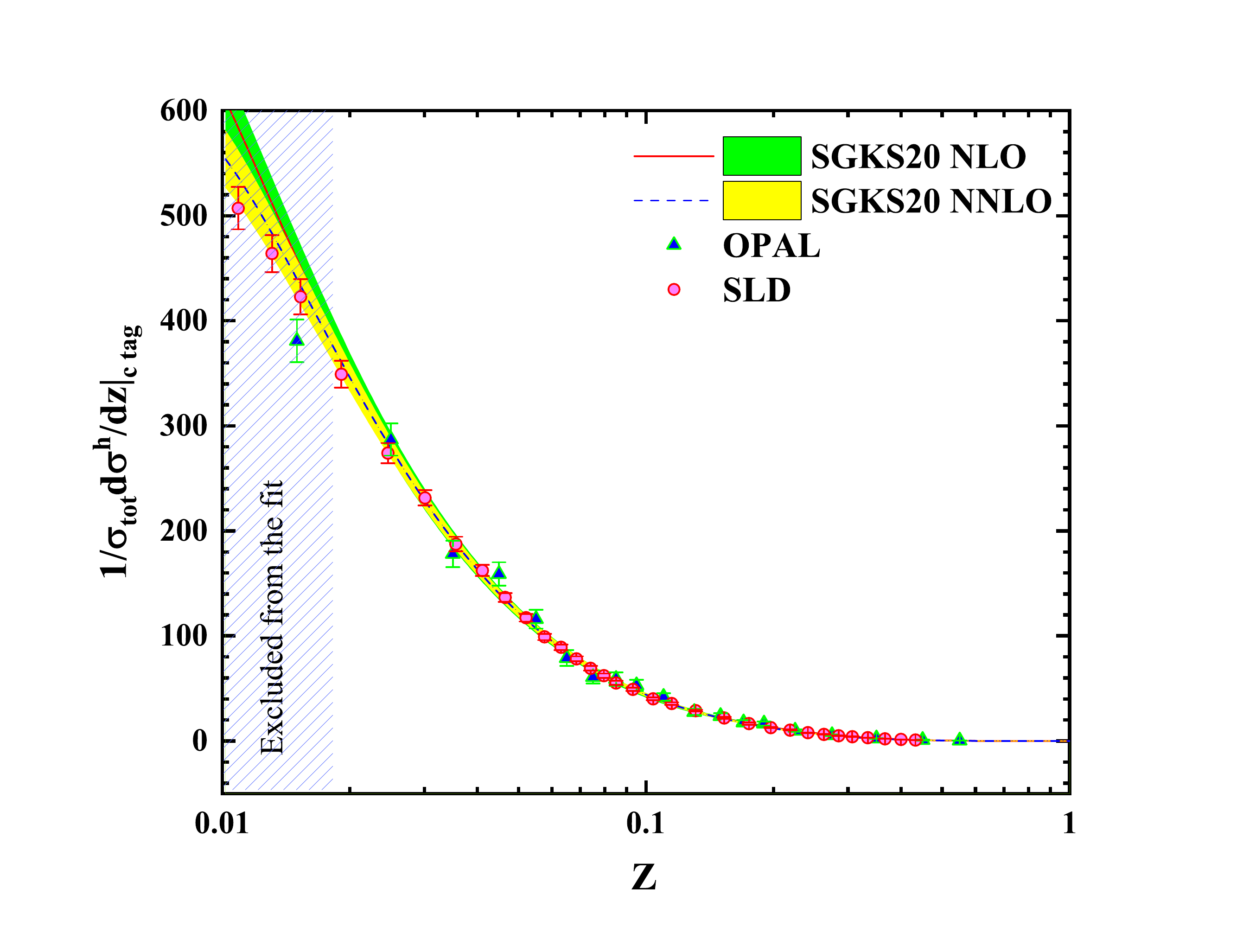}} 	 	
\resizebox{0.480\textwidth}{!}{\includegraphics{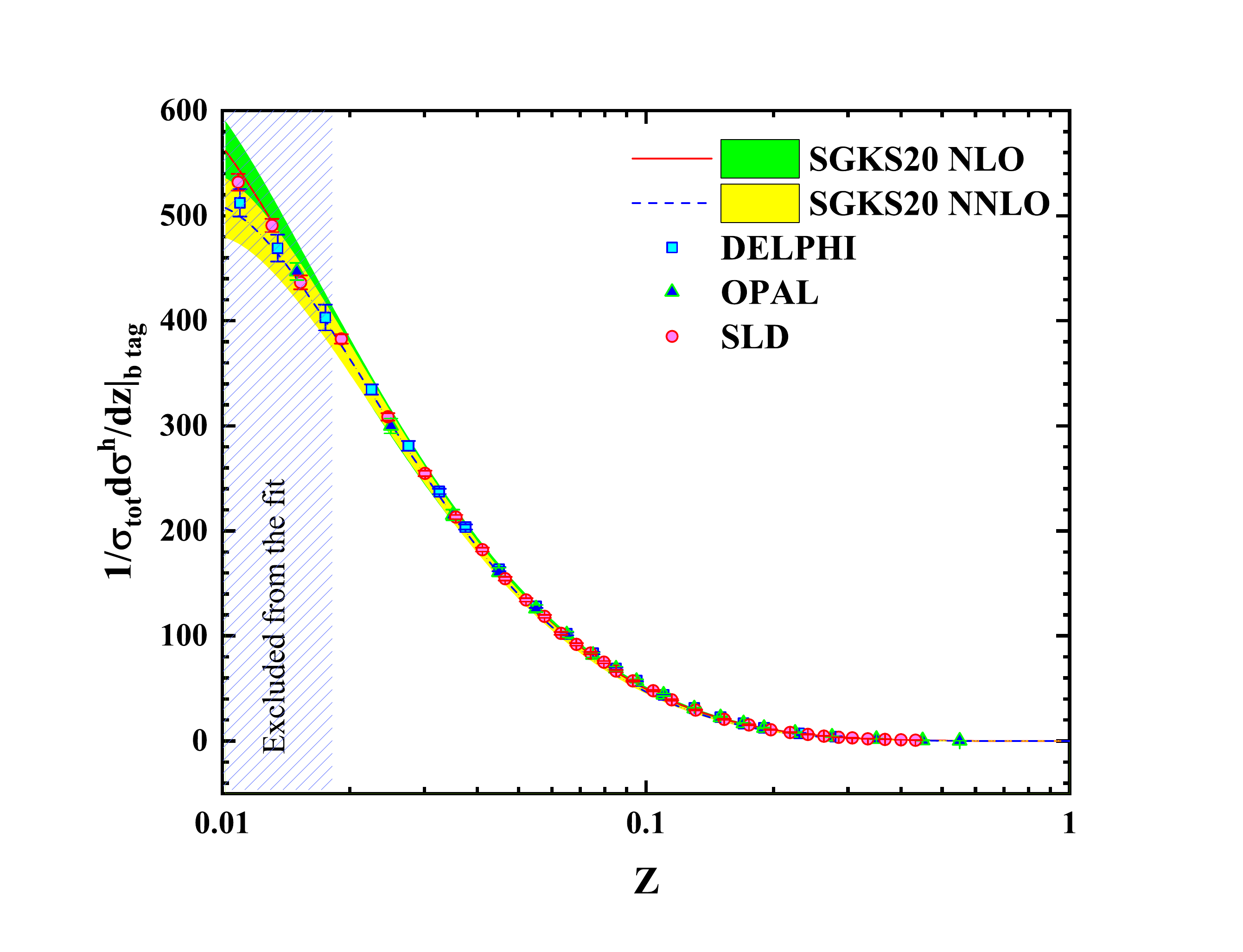}} 	
\begin{center}
\caption{ \small 
			NLO and NNLO theory predictions for the normalized SIA cross sections 
			of unidentified light charged hadrons in comparison with the total 
			inclusive~\cite{Buskulic:1994ft,Buskulic:1995aw,Abreu:1998vq, 
				Akers:1994ez,Ackerstaff:1998hz}, 
			light~\cite{Abreu:1998vq,Akers:1994ez,Ackerstaff:1998hz, 
				Abe:2003iy}, $c$-tagged~\cite{Akers:1994ez,Ackerstaff:1998hz, 
				Abe:2003iy} and $b$-tagged~\cite{Abreu:1998vq,Akers:1994ez,
				Ackerstaff:1998hz,Abe:2003iy} SIA cross section measurements from 
			the {\tt ALEPH}, {\tt DELPHI}, {\tt OPAL} and {\tt SLD} experiments. 
			The green (NLO) and yellow (NNLO) shaded bands correspond to 
			uncertainty estimates based on the Hessian approach with 
			$\Delta \chi^2 = 1$. 
} 
\label{fig:SIA}
\end{center}
\end{figure*}

Having at hand the {\tt SGKS20} NLO and NNLO light charged hadron 
FFs, we are now able to compare the analyzed data against the theory 
predictions for the normalized SIA cross sections. In 
Fig.~\ref{fig:SIA}, our theory predictions 
are compared to the total SIA cross section measurements for 
inclusive~\cite{Buskulic:1994ft,Buskulic:1995aw,Abreu:1998vq,
Akers:1994ez,Ackerstaff:1998hz}, 
light~\cite{Abreu:1998vq,Akers:1994ez,Ackerstaff:1998hz,Abe:2003iy}, 
$c$- tagged~\cite{Akers:1994ez,Ackerstaff:1998hz,Abe:2003iy} and 
$b$-tagged~\cite{Abreu:1998vq,Akers:1994ez,Ackerstaff:1998hz,Abe:2003iy}   
unidentified light charged hadron $(h^\pm)$ from {\tt ALEPH}, 
{\tt DELPHI}, {\tt OPAL} and {\tt SLD} experiments.
In general, the agreement between data and theory is excellent. 
In addition, we observe that our NNLO results show a better agreement 
with the SIA data, especially for the total inclusive, $c$-tagged and 
light charged hadron cross sections at small values of $z$. One can 
also see that the error bands for the NLO and NNLO theory predictions 
are very similar, except for the case of $c$-tagged cross sections 
where the NNLO predictions show smaller uncertainties.

\begin{figure*}[htb]
\includegraphics[width=0.64\linewidth,trim=40 10 35 20,clip]{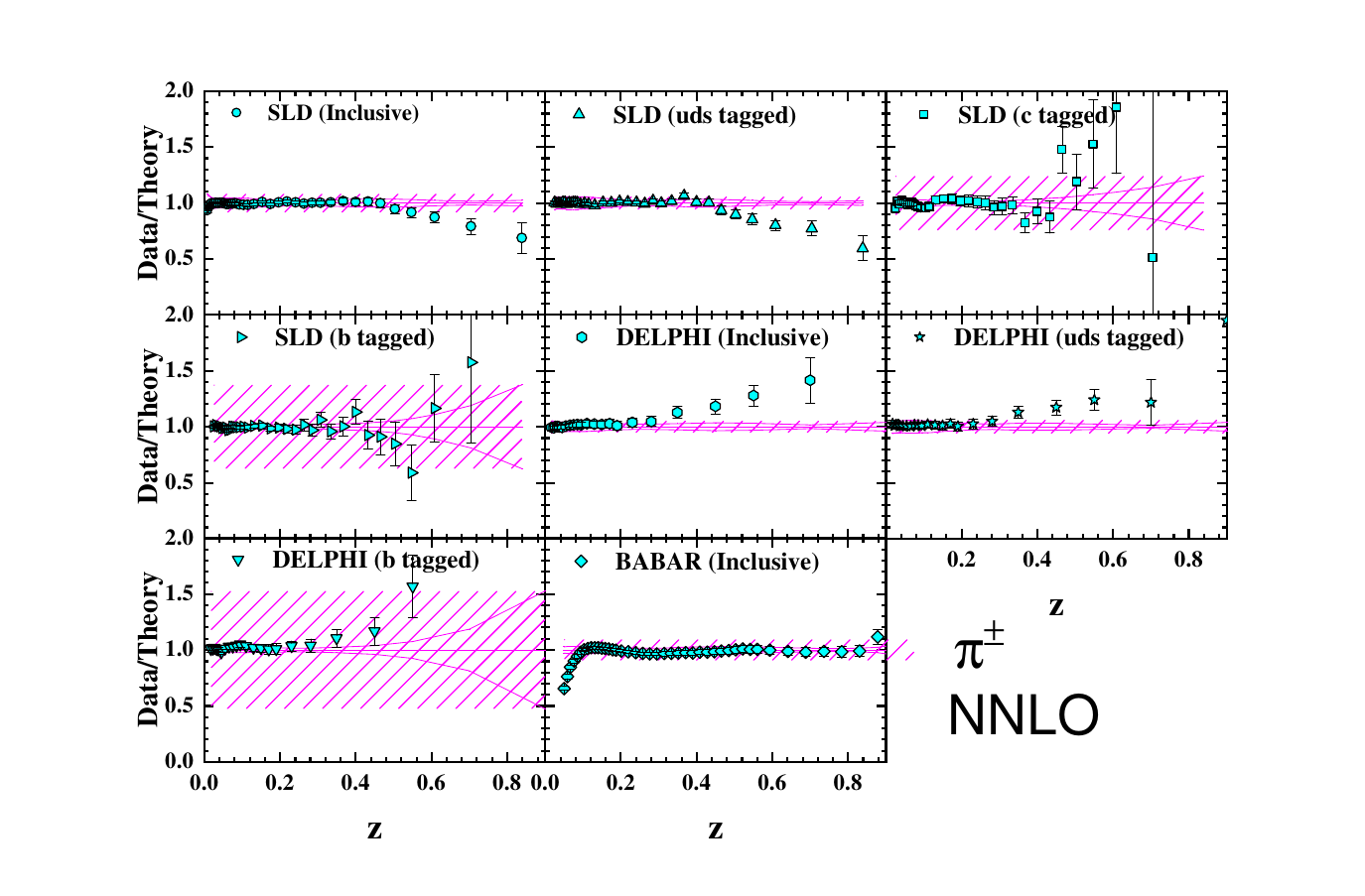}		
\begin{center}
\caption{ \small 
  The data/theory ratio for the charged pion ($\pi^\pm$) production 
  data from {\tt SLD}~\cite{Abe:2003iy}, 
  {\tt DELPHI}~\cite{Abreu:1998vq} and {\tt BABAR}~\cite{Lees:2013rqd} 
  experiments included in the {\tt SGKS20} fit. Our theoretical 
  predictions are computed at NNLO accuracy with our best-fit NNLO FFs. }
\label{fig:Pion-ratio}
\end{center}
\end{figure*}

\begin{figure*}[htb]
\includegraphics[width=0.64\linewidth,trim=40 10 35 20,clip]{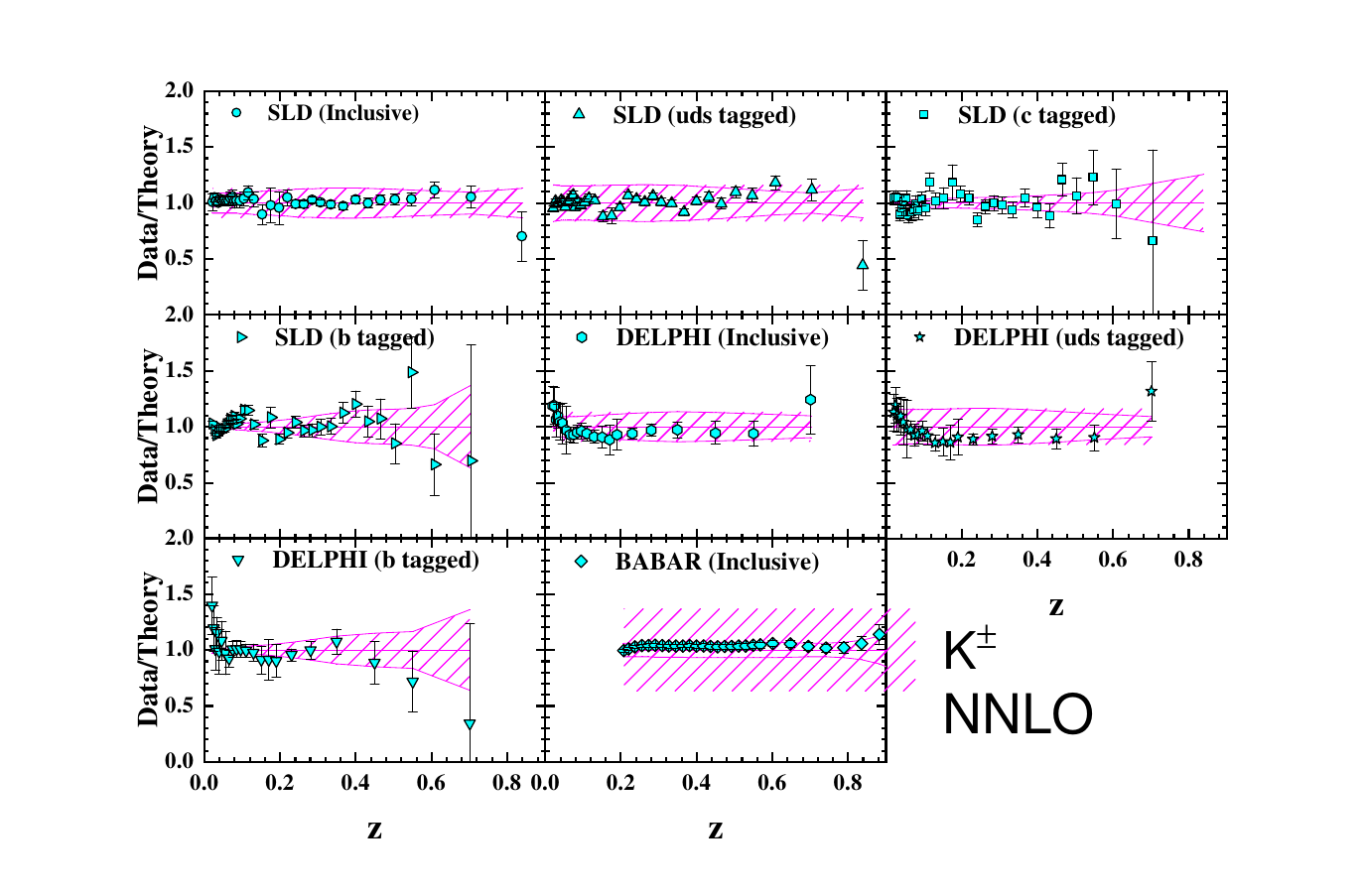}  		
\begin{center}
\caption{ \small 
  Same as Fig.~\ref{fig:Pion-ratio} but for charged kaons ($K^\pm$). 
  }
\label{fig:kaon-ratio}
\end{center}
\end{figure*}

\begin{figure*}[htb]
\includegraphics[width=0.64\linewidth,trim=40 10 35 0,clip]{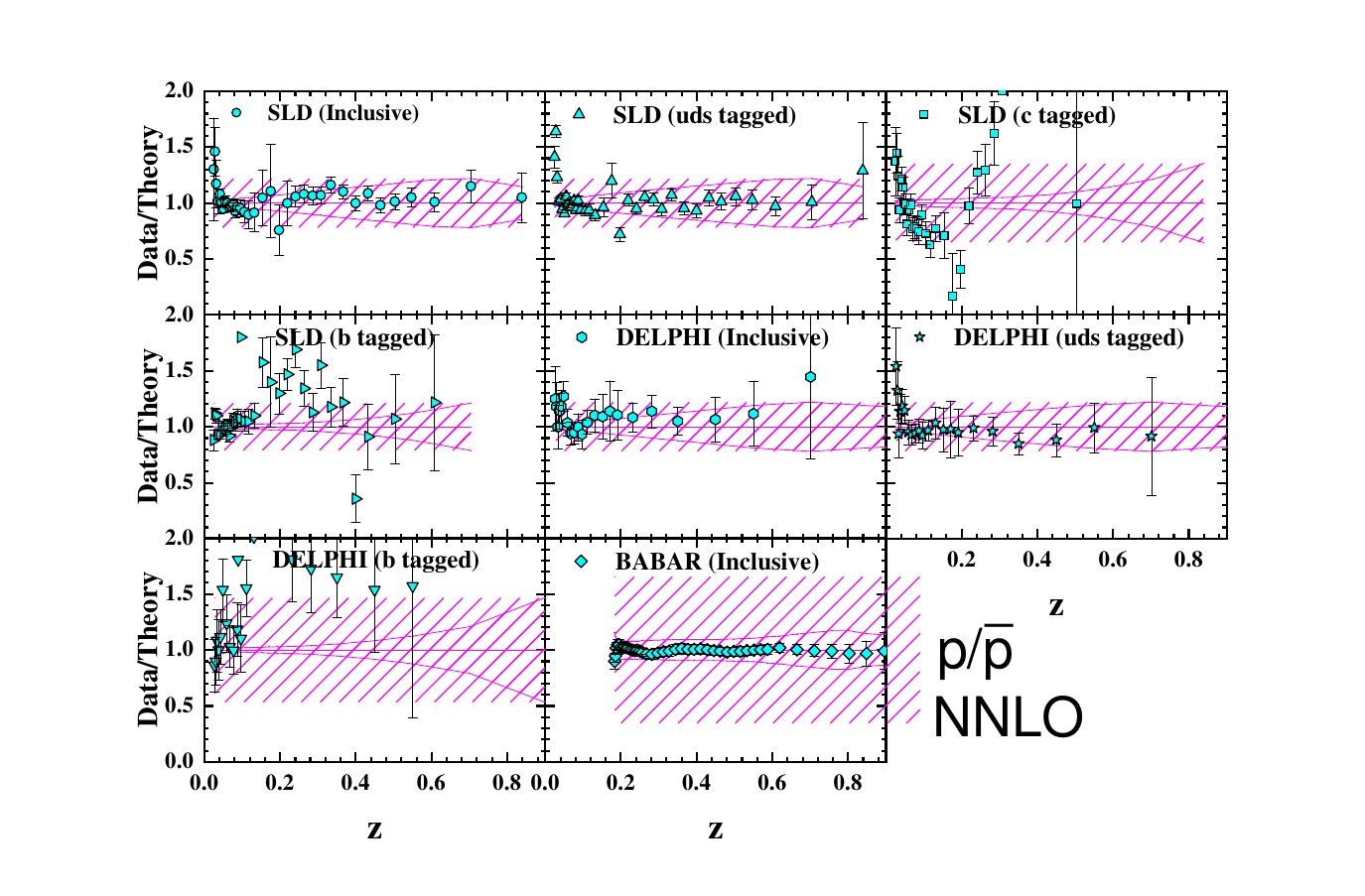} 		
\begin{center}
\caption{ \small 
  Same as Fig.~\ref{fig:Pion-ratio} but for protons/antiprotons, 
  ($p /\bar p$). 
  }
\label{fig:Proton-ratio}
\end{center}
\end{figure*}

We also present a comparison of the charged pion, kaon and 
proton/antiproton data used in this analysis with the
corresponding theoretical predictions obtained using our NNLO FFs. 
In Figs.~\ref{fig:Pion-ratio}, \ref{fig:kaon-ratio} and 
\ref{fig:Proton-ratio}, data over theory ratios are displayed for the 
{\tt SLD}~\cite{Abe:2003iy}, {\tt DELPHI}~\cite{Abreu:1998vq} and 
{\tt BABAR}~\cite{Lees:2013rqd} data for charged pion ($\pi^\pm$), 
charged kaon ($K^\pm$), and proton/antiproton ($p /\bar p$) 
production in SIA.

For the case of pion production, Fig.~\ref{fig:Pion-ratio}, overall 
good agreement between measurements and the NNLO theory predictions 
is found for most of the data points, except for the large-$z$ region 
of some experiments. The uncertainties of our theory predictions 
are getting large in this region for the case of {\tt  SLD} and 
{\tt DELPHI} heavy quark production.

The comparison for charged kaons is presented in 
Fig.~\ref{fig:kaon-ratio}. We notice that for some data the agreement 
is good, in particular for the {\tt SLD} and {\tt BABAR} experiments, 
while for {\tt DELPHI} we see some deviations in the small-$z$ region. 
As one can see, the experimental data points for all data sets fluctuate 
inside the error bands of the theory predictions.

Finally, we display in Fig.~\ref{fig:Proton-ratio} the data/theory 
ratios for proton/antiproton ($p /\bar p$) production for all 
experimental data analyzed in this work. One can see that for $c$- 
and $b$-tagged data the agreement is poor, but the comparison between 
our predictions and the total inclusive and $uds$-tagged data is 
reasonable. Deviations appear specifically for almost all experiments 
in the small-$z$ region, except for the case of inclusive measurements 
from the {\tt BABAR} experiment. For the inclusive measurements of 
{\tt SLD}, {\tt DELPHI} and {\tt BABAR}, the agreement is acceptable 
in the medium-to-large range of $z$-values. The same conclusion can 
be made for the $uds$-tagged data from the {\tt SLD} and {\tt DELPHI} 
experiments.

%
\section{Summary and conclusions} 
\label{sec:conclusion}
%

The main goal of the current study is to present a set of FFs, 
called {\tt SGKS20}, for light charged hadron ($\pi^\pm$, $K^\pm$, 
$p/\bar{p}$) production. These FFs are obtained in a simultaneous 
fit and we include both identified and unidentified light charged 
hadron data taken from electron-positron annihilation. We included 
finite-hadron mass corrections which are significant for small $z$ 
and small $\sqrt{s}$. For FFs which involve heavy quarks, we adopted 
the zero-mass variable-flavour-number scheme. As a third improvement, 
the \textit{residual} light hadrons contributions have been included 
in our fit for unidentified light hadrons. We have shown that this 
approach improves the total $\chi^2$ at both NLO and NNLO accuracy 
and also reduces the uncertainties for the FFs of light hadrons. Our 
results show that the inclusion of higher-order QCD corrections 
helped to obtain a much better agreement of data with theory. 
Finally, we compared our pion, kaon and proton FFs with the one 
recently extracted by the {\tt NNFF1.0} Collaboration.

The most important limitation of the present analysis is related 
to the fact that we have included data from SIA measurements only. 
The precise data from proton-(anti)proton ($pp$) collisions, which 
cover a wide range in energy and momentum fractions, contain vital 
information about FFs, especially for the gluon FF, and also are 
sensitive to different partonic combinations~\cite{Bertone:2018ecm}. 
These measurements include {\tt CDF}~\cite{Abe:1988yu,Aaltonen:2009ne} 
experiment at the Tevatron, {\tt STAR}~\cite{Adamczyk:2013yvv} and 
{\tt PHENIX}~\cite{Adare:2007dg} at RHIC, and 
{\tt CMS}~\cite{Chatrchyan:2011av,CMS:2012aa} and 
{\tt ALICE}~\cite{Abelev:2013ala} experiments at the LHC. 
It is expected that the inclusion of these data will lead to much 
better constrained FFs. Hence, it will be interesting to repeat this 
analysis by considering the SIDIS data as well as hadron collider data 
which could provide a flavor separation between quark and antiquark 
FFs and also the gluon FF. In addition, a future study investigating 
the improvements of description of the data at low center-of-mass 
energy due to the effect arising from heavy quarks mass corrections 
would be very interesting.

The FF parametrizations at NLO and NNLO for identified light charged 
hadron, $zD^{H^\pm} \, (H^\pm = \pi^\pm, K^\pm, p/\bar{p})$, presented 
in this study are available in the standard {\tt LHAPDF} 
format~\cite{Buckley:2014ana} from the authors upon request.
\\
\\

%
\begin{acknowledgments}
%

The authors are thankful to Valerio Bertone and Vadim Guzey for many 
helpful discussions and comments.
The Authors thank the School of Particles and Accelerators, Institute 
for Research in Fundamental Sciences (IPM) for financial support of 
this project.
Hamzeh Khanpour also is thankful to the University of Science and 
Technology of Mazandaran for financial support provided for this 
research.

\end{acknowledgments}

\clearpage


\end{document}